# EUROPEAN LONGITUDE PRIZES
# I. LONGITUDE DETERMINATION IN THE SPANISH EMPIRE

**Richard de Grijs**
*Department of Physics and Astronomy, Macquarie University,*
*Balaclava Road, Sydney, NSW 2109, Australia*
Email: richard.de-grijs@mq.edu.au

**Abstract:** Following Columbus' voyages to the Americas, Castilian (Spanish) and Portuguese rulers engaged in heated geopolitical competition, which was eventually reconciled through a number of treaties that divided the world into two unequal hemispheres. However, the early-sixteenth-century papal demarcation line was poorly defined. Expressed in degrees with respect to a vague reference location, determination of longitude at sea became crucial in the nations' quest for expanding spheres of influence. In Spain, King Philip II and his son, Philip III, announced generous rewards for those whose solutions to the longitude problem performed well in sea trials and which were suitable for practical implementation. The potential reward generated significant interest from scientist-scholars and opportunists alike. The solutions proposed and the subset taken to sea provided important physical insights that still resonate today. None of the numerous approaches based on compass readings ('magnetic declination') passed the exacting sea trials, but the brightest sixteenth-century minds already anticipated that lunar distances and the use of marine timepieces would eventually enable more precise navigation. With most emphasis in the English-language literature focused on longitude solutions developed in Britain, France and the Low Countries, the earlier yet groundbreaking Spanish efforts have, undeservedly, largely been forgotten. Yet, they provided a firm basis for the development of an enormous 'cottage industry' that lasted until the end of the eighteenth century.

**Keywords:** longitude determination, Spanish longitude prizes, terrestrial magnetism, lunar distances, Jupiter's satellites, demarcation lines

## 1 SOUTHERN EUROPEAN RIVALRY

By the end of the fifteenth century, Christopher Columbus (see Figure 1[1]) still practiced 'dead reckoning'—maintaining a predetermined (westerly) compass heading at a fixed latitude—on his voyages across the Atlantic. He relied largely on Claudius Ptolemy's seminal world atlas, *Geographia*, the most complete collection of geographic knowledge in the second century CE. However, Ptolemy's maps were unsuitable for precision navigation. They were based on erroneous conversions from distances into degrees and *vice versa*, for instance leading to significantly exaggerated representations of the Mediterranean region. Throughout the late-medieval transition period and the Renaissance, these problematic cartographic renditions were gradually eliminated as improved measurements of terrestrial features and distances became available (de Grijs, 2017: Ch. 2).

Politically, determination of one's accurate position at sea and in the newly discovered territories of the West Indies became a key issue in the ongoing rivalry between Castile (for simplicity henceforth identified as Spain[2]) and Portugal, major and arguably the dominant European maritime powers in the fifteenth and sixteenth centuries. On 4 September 1479, the Catholic Sovereigns of Aragon and Castile, King Ferdinand II and Queen Isabella I (see Figure 2[3]), had signed the *Treaty of Alcáçovas*[4] with King Afonso (Alfonso) V of Portugal and the Algarves, 'the African', and his son, 'the Perfect' Prince João (John) II of Portugal, to end the War of the Castilian Succession (1475–1479). Ferdinand and Isabella had been victorious on land, whereas the Portuguese had established their hegemony at sea (Diffie and Winius, 1985; Newitt, 2005).

The *Treaty* divided the Atlantic into Spanish and Portuguese spheres of influence, except for the Canary Islands. Confirmed on 21 June 1481 through Pope Sixtus IV's bull *Æterni regis* (*Eternal king's*), the *Treaty of Alcáçovas* granted the Canary Islands to Spain. All lands south of the Canaries acquired by Christian powers—in the South Atlantic (Madeira, the Azores, Cape Verde), along the African coast (Guinea), and eastwards to the Indies—were granted to Portugal:

> … the aforesaid King and Queen of Castile, Aragon and Sicily … promised henceforth and forever that neither directly nor indirectly, neither secretly nor publicly, nor by their heirs and successors, will they disturb, trouble or molest, in fact or in law, in court or out of court, the said

King and Prince of Portugal or the future sovereigns of Portugal or their kingdoms in the status of possession or quasi-possession which they hold over all the trade, lands and barter of Guinea and the Gold Coast, or over any other islands, shores, sea coasts or lands, discovered or to be discovered, found or to be found, or over the islands of Madeira, Porto Santo and Desertas, or over all the islands called the Azores, that is, Ancipitrum, and Flores Island, nor over the islands of Cape Verde, nor over the islands already discovered, nor over whatever islands shall be found or acquired from beyond the Canaries and on this side of and in the vicinity of Guinea, so that whatever has been or shall be found and acquired further in the said limits shall belong to the said King and Prince of Portugal and to their kingdoms, excepting only the Canary Islands, [namely] Lanzarote, La Palma, Fuerteventura, La Gomera, Ferro [El Hierro], La Graciosa, Gran Canaria, Tenerife and all other Canary Islands, acquired or to be acquired, which remain the possession of the kingdoms of Castile. (Davenport, 1917)

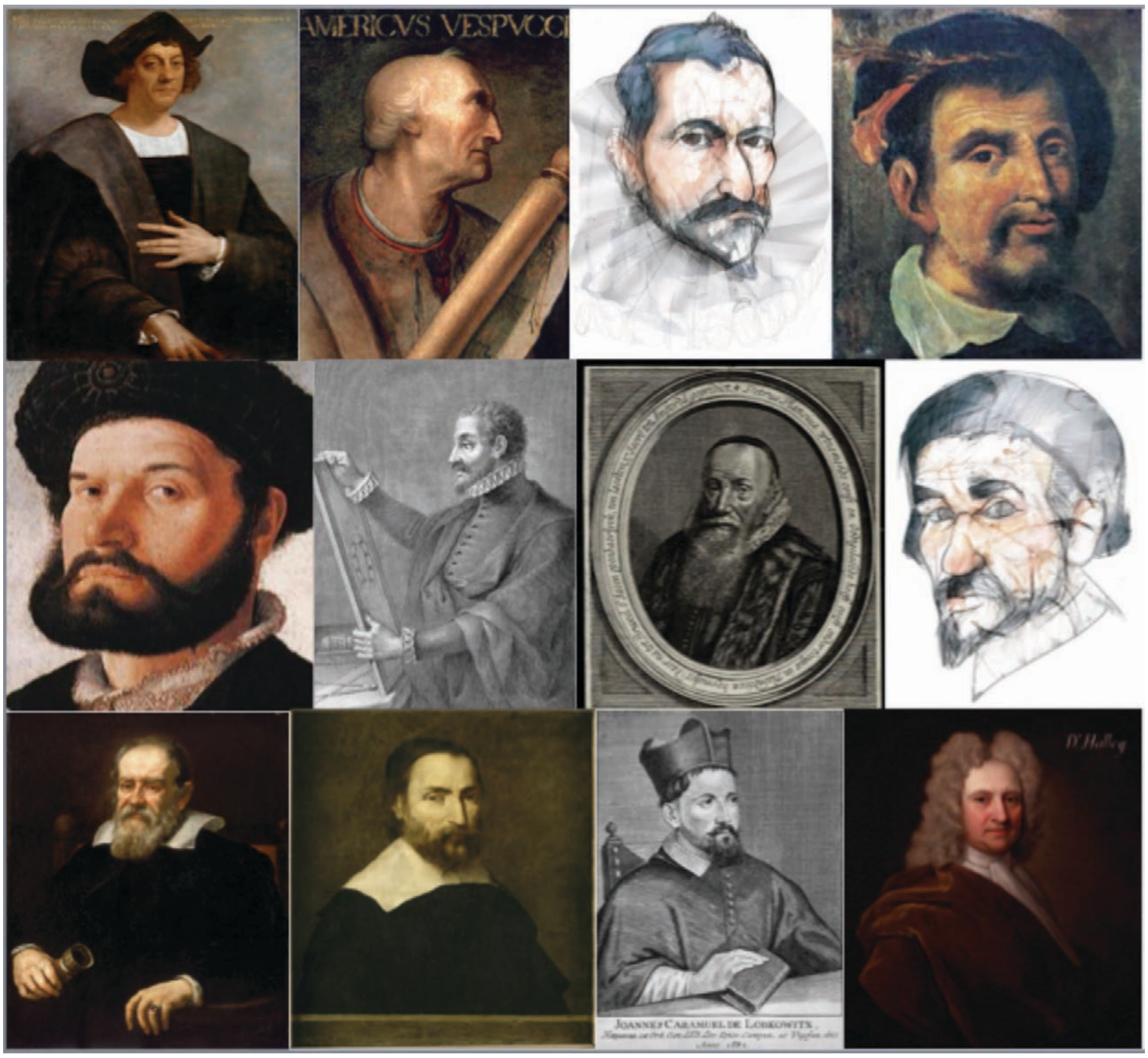

**Figure 1**: Portraits of the main characters driving the innovations described in this article, both scientist-scholars and opportunistic ‘projectors’, ordered from left to right and from top to bottom by date or birth. Individuals depicted include *(i)* Christopher Columbus (1451–1506); *(ii)* Vespucci (1454–1512); *(iii)* Guillén (1487/1492–1561); *(iv)* and *(v)* Ferdinand Columbus (1488–1539); *(vi)* de Herrera (1530–1593); *(vii)* Plancius (1552–1622); *(viii)* Ayanz (1553–1613); *(ix)* Galileo (1564–1642); *(x)* de Peiresc (1580–1637); *(xi)* Caramuel (1606–1682); *(xii)* Halley (1656–1742). *Figure credits*: Wikimedia Commons, except for *(vi)*: https://alchetron.com/Ferdinand-Columbus. All images are in the public domain, except for *(iii)*, *(vi)* and *(viii)*: Creative Commons Attribution-Share Alike 4.0 International license.

In effect, the *Æterni regis* bull confirmed and superseded Pope Nicholas V’s earlier *Dum Diversas* (*Until different*, 1452), *Romanus Pontifex* (*The Roman Pontiff*, 1455) and *Inter Caetera* (*Among other [works]*, 1456) bulls. Through his *Æterni regis* bull, Pope Sixtus IV formally recognised Portugal’s territorial acquisitions along the West African coast. Papal law, eventually converted into common law, thus provided the legal basis of Columbus’ authority as viceregal representative of the Spanish Crown in the New World. The body of fifteenth-century papal bulls also re-emphasised the pope’s authority, as Christ’s representative in both spiritual and temporal matters, over the competing nations' interests. The legal basis of the papal bulls themselves traces back to a fourth-century CE edict issued by the Roman Emperor, Constantine the Great, to Pope Sylvester, conferring on the latter the authority over Rome and the western expanses of the Roman Empire.

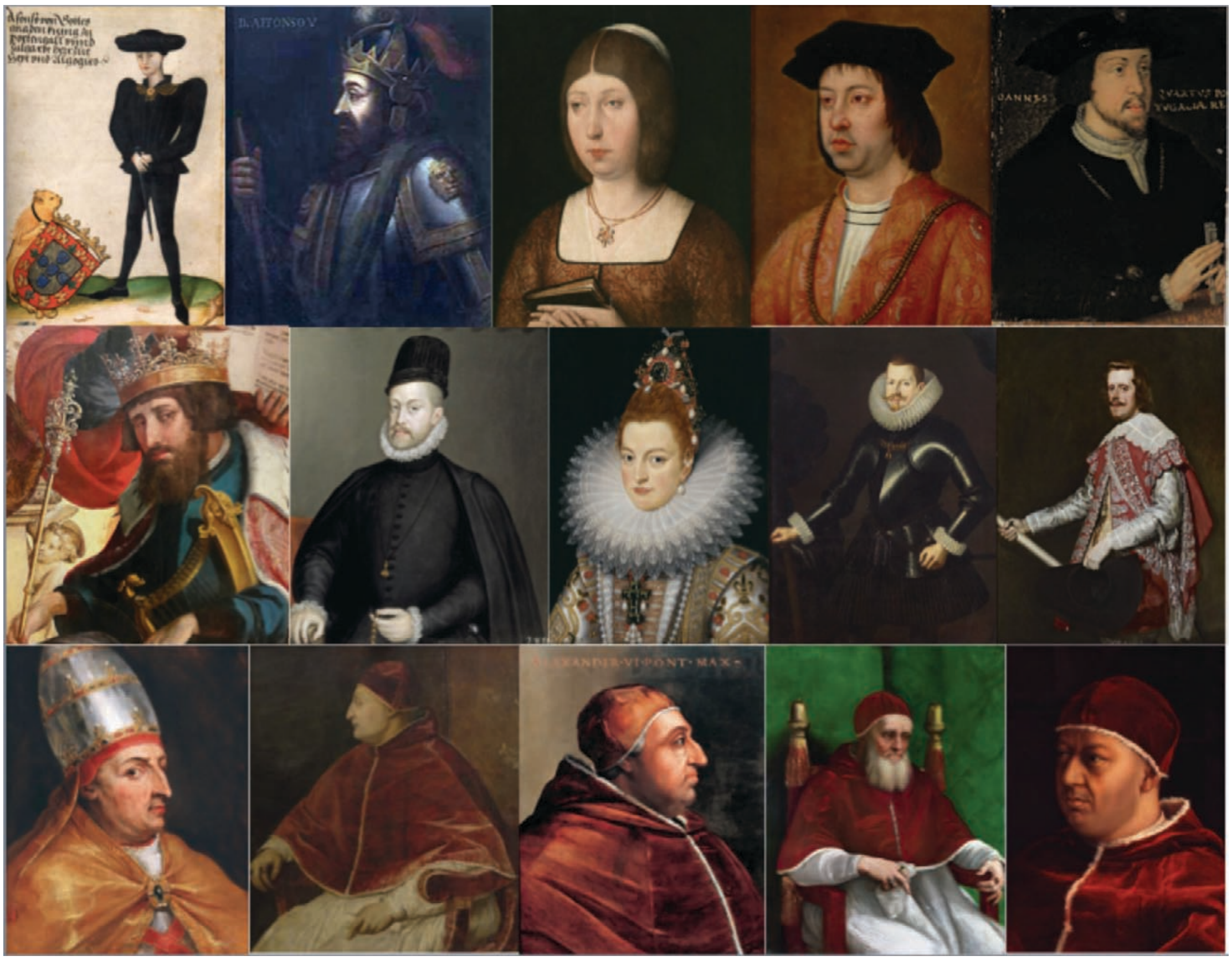

**Figure 2**: As Figure 1 but for the royalty (top and middle rows) and popes (bottom row) playing central roles in our narrative. *(i)* King Alfonso V (1432–1481); *(ii)* Queen Isabella I (1451–1504); *(iii)* King Ferdinand II (1452–1516); *(iv)* Prince/King John II (1455–1495); *(v)* King Emmanuel I (1469–1521); *(vi)* King Philip II (1527–1598); *(vii)* Infanta Isabella Clara Eugenia (1566–1633); *(viii)* King Philip III (1578–1621); *(ix)* King Philip IV (1605–1665); *(x)* Pope Nicholas V (1397–1455); *(xi)* Pope Sixtus IV (1414–1484); *(xii)* Pope Alexander VI (1431–1503); *(xiii)* Pope Julius II (14443–1513); *(xiv)* Pope Julius II (14443–1513); *(xv)* Pope Leo X (1475–1521). *Figure credits*: Wikimedia Commons (public domain).

Columbus' first voyage of discovery across the Atlantic Ocean was in direct violation of the *Treaty of Alcáçovas.* Upon his return in 1493, he first arrived in Lisbon, Portugal, where he impressed by-then-King John II with his new discoveries. Mindful of the *Treaty*, the Portuguese king announced to the Spanish sovereigns that all lands discovered by Columbus—who had actually sailed under Castilian patronage—did, in fact, belong to Portugal. King Ferdinand and Queen Isabella opted for a diplomatic resolution, which eventually resulted in the *Treaty of Tordesillas* (7 June 1494; see Figure 3).

## 2 A CONTROVERSIAL PAPAL DEMARCATION LINE

On 26 September 1493 Pope Alexander VI issued an updated bull, *Dudum siquidem* (*Extension of the Apostolic Grant and Donation of the Indies*), which awarded all mainlands and islands, "at one time or even yet belonged to India" to the Catholic monarchs, no matter where they might be located geographically:

> A short while ago … we gave, conveyed and assigned forever to you and your heirs and successors … all islands and mainlands whatsoever, discovered and to be discovered, toward the west and south, that were not under the actual temporal dominion of any Christian lords. … But since it may happen that your envoys and captains, or vassals, while voyaging toward the west or south, might bring their ships to land in eastern regions and there discover islands and mainlands

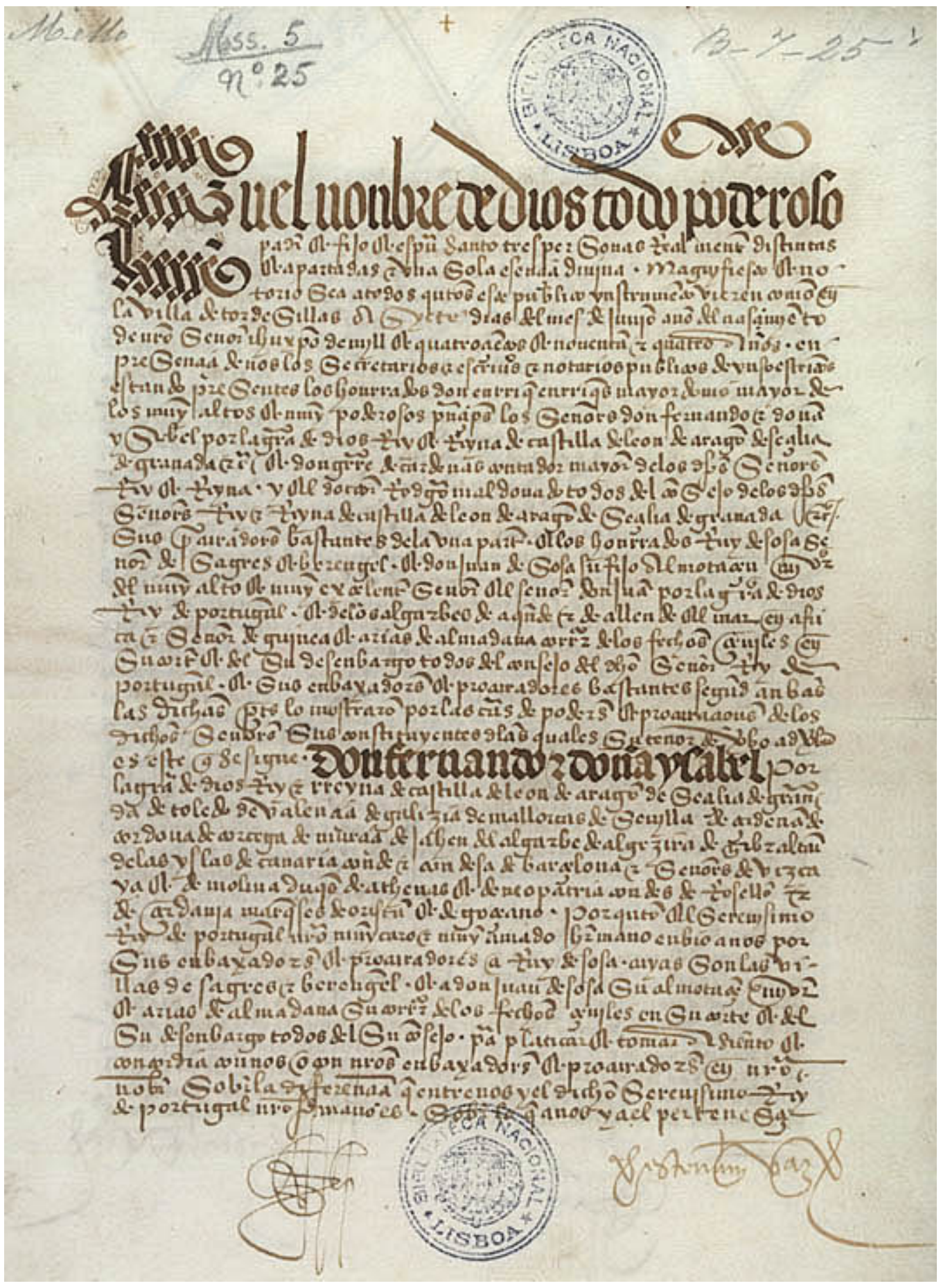

**Figure 3**: First page of the *Treaty of Tordesillas*, 1494. (Biblioteca Nacional de Lisboa via Wikimedia Commons; public domain)

that belonged or belong to India, … we do in like manner amplify and extend our aforesaid gift, grant, assignment and letters, with all and singular the clauses contained in the said letters, to all islands and mainlands whatsoever, found and to be found, discovered and to be discovered, that are or may be or may seem to be in the route of navigation or travel towards the west or south, whether they be in western parts, or in the regions of the south and east and of India. We grant to you and your aforesaid heirs and successors full and free power through your own authority, exercised through yourselves or through another or others, freely to take corporal possession of the said islands and countries and to hold them forever, …

In response, and disturbed by the explicit reference to India, a major territorial growth area for the Portuguese, the king of Portugal bypassed the pope and opened direct negotiations with representatives of Ferdinand and Isabella. King John II's aim was that the papal demarcation line be moved to the west. He agreed to adopt the revised *Inter Caetera* bull of 1493, that is, the second version of this decree, as starting point for the negotiations. As a result, the meridian line was moved 270 leagues to the west, which meant that eastern Brazil now became part of the Portuguese sphere of influence. Competition between both maritime nations intensified:

> … both sides must have known that so vague a boundary could not be accurately fixed, and each thought that the other was deceived, [concluding that it was a] diplomatic triumph for Portugal, confirming to the Portuguese not only the true route to India, but most of the South Atlantic. (Parry, 1973)

Pope Julius II, the 'Fearsome' or 'Warrior' Pope, subsequently sanctioned the updated *Treaty* by means of a bull issued on 24 January 1506, *Ea quae pro bono pacis* (*For the sake of peace*):

> A request recently addressed to us on the part of our very dear son in Christ, Emmanuel, the illustrious, King of Portugal and of the Algarves, stated that inasmuch as some time ago the permission was granted by the Apostolic See to John, of illustrious memory, King of Portugal and the Algarves, to the effect that the said John and any King of Portugal and of the Algarves for the time being, should be permitted to navigate the ocean sea, or seek out the islands, ports and mainlands lying within the said sea, and to retain those found for himself, and to all others it was forbidden under penalty of excommunication, … from presuming to navigate the sea in this way against the will of the aforesaid king, or to occupy the islands and places found there; and inasmuch as between the aforesaid King John, on the one part, and our very dear son in Christ, Ferdinand, at that time the illustrious King of Aragon, Castile and León, on the other part, in regard to certain islands called Las Antillas, which had been discovered and occupied by the aforesaid king, … came to a certain honourable agreement, convention and compact, whereby, among other things, they resolved that the Kings of Portugal and the Algarves should have the right to navigate the said sea within certain specified limits and seek out and take possession of newly discovered islands and that the Kings … of Castile and Leon should have the same right

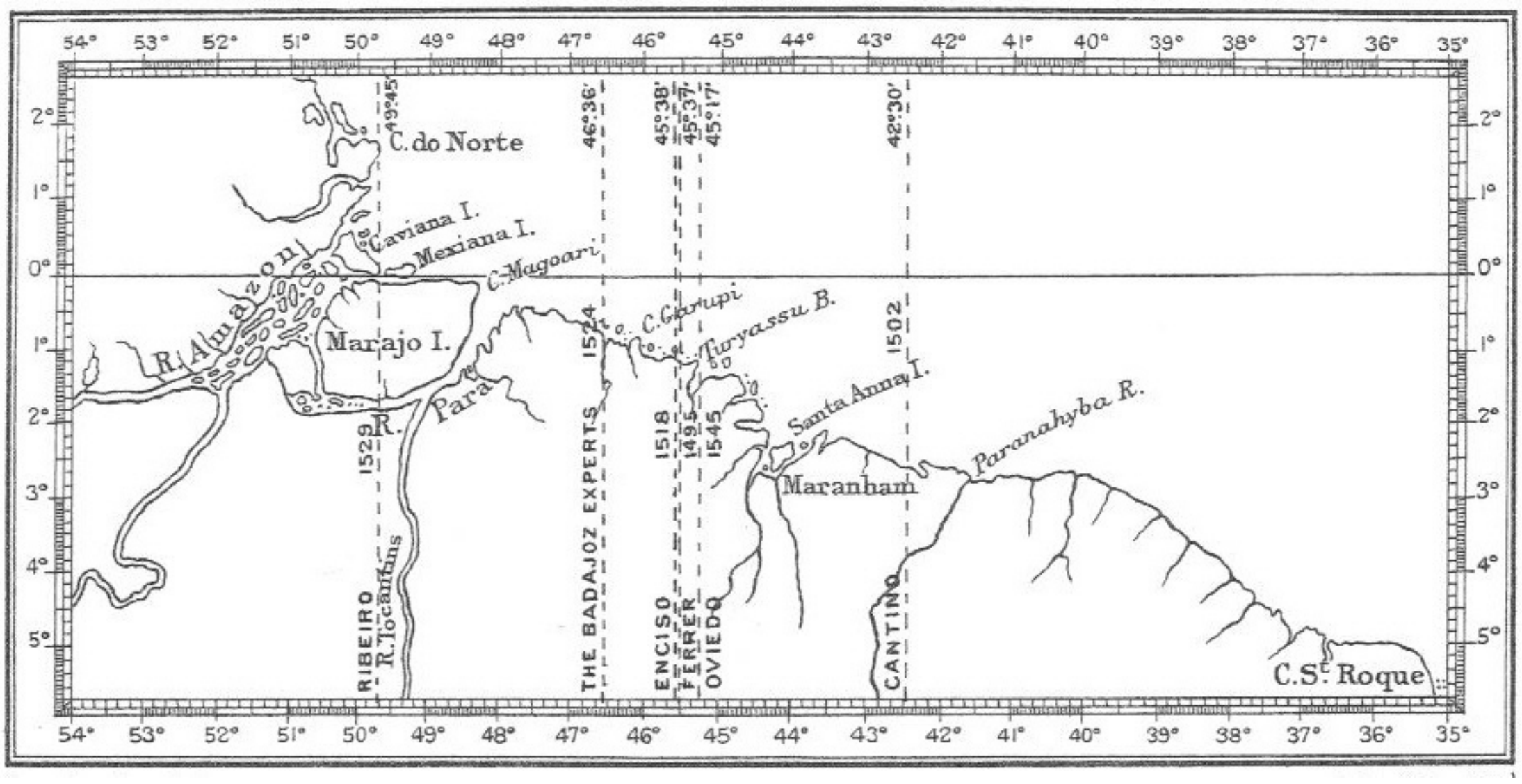

**Figure 4**: Early demarcation lines (1495–1545) associated with the *Treaty of Tordesillas*. (Harrisse, H., 1897. *The Diplomatic History of America: Its first chapter 1452–1493–1494*. London, Stevens. Frontispiece; public domain)

> within certain other specified limits … Wherefore the aforesaid King Emmanuel has humbly besought us to deign to add the authority of the apostolic confirmation to the aforesaid agreement, convention and compact for the purpose of establishing them more firmly … (Davenport, 1917)

King Manuel I (Emmanuel), 'the Fortunate', had zealously worked on expansion of the Catholic faith in Morocco, Guinea and India. Pope Julius II was, hence, impressed by the Portuguese king's efforts to proselytise and therefore well-disposed towards his requests for an expanded sphere of influence. Pope Julius II's successor, Pope Leo X, treated King Emmanuel's requests and ambassadorial representations similarly favourably. Indeed, a decade later and at the insistence of Portugal (which attempted to limit Spain's expansion in Asia), Pope Leo X issued his bull *Praecelsae devotionis* (*Devotion of the Nobles*, 1514), pronouncing that the line of demarcation applied to the Atlantic Ocean only:

> … we newly grant everything, all and singular, contained in the aforesaid letters, and all other empires, kingdoms, principalities, duchies, provinces, lands, cities, towns, forts, lordships, islands, harbours, seas, coasts and all property, real and personal, wherever existing, also all unfrequented places, recovered, discovered, found and acquired from the aforesaid infidels, by the said King Emmanuel and his predecessors, or in the future to be recovered, acquired, discovered and found by the said King Emmanuel and his successors, both from Cape Bojador [a headland on the coast of Western Sahara, south of the Canary Islands] and Não [a cape on the southern coast of Morocco] to the Indies, and in any place or region whatsoever, even although perchance unknown to us at present … (Parry, 1973: 202)

Be that as it may, in the early sixteenth century the key problem of the determination of one's accurate position at sea, specifically of one's longitude, had not yet been resolved (e.g., de Grijs, 2017). Irrespective of this crucial shortcoming, the original *Treaty of Tordesillas*, as well as the updated version of 1506, identified the demarcation line in terms of the number of leagues West of the Azores and Cape Verde Islands—that is, in the open ocean, where geographic position measurements were all but impossible using contemporary tools. The *Treaty* remained vague on detail as regards which of the islands should be adopted as point of reference, or exactly which measurement of the league[5] to employ. This thus left the precise longitude of the demarcation line, a reference measured in degrees, undefined. It was specified that these issues had to be resolved through a joint voyage, which however never materialised. Instead, at least five different opinions as to the exact location of the line of demarcation surfaced between 1495 and 1524 (see Figure 4 and, for a modern equivalent, Figure 5):

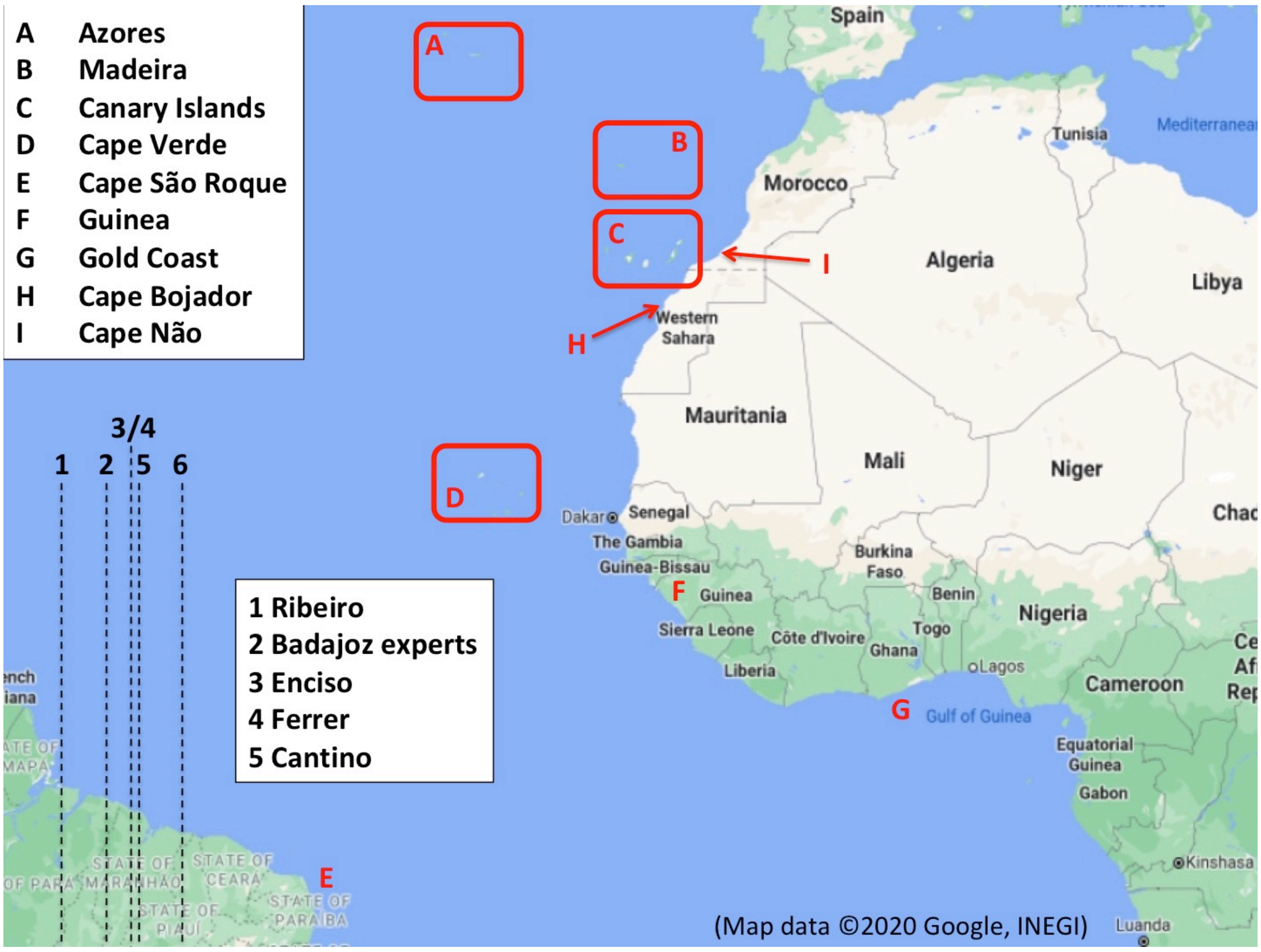

**Figure 5**: Modern representation of the early demarcation lines (numbered vertical dotted lines) and the most important geographic locations in the mid-Atlantic Ocean. (Map data © 2020 Google, INEGI; permitted use)

- At the request of Ferdinand and Isabella, the cosmographer Jaume (Jaime) Ferrer de Blanes (see Figure 6[6]) concluded in 1495 that the line was located 18° West of the central island of the Cape Verde archipelago, identified with Fogo. The nineteenth-century historian Henry Harrisse (1897: 91–97, 178–190) determined that the de Blanes line of demarcation corresponded to a longitude, in modern units, of 47°37′ W, 2276.5 km West of Fogo.

- On the Portuguese *Cantino* planisphere of 1502 (see Figure 7), the line of demarcation is located between Cape São Roque at the northeastern tip of Brazil and the Amazon river estuary. Harrisse (1897: 100–102, 190–192) located the line at 42°30′ W in modern units.

- In 1518, Martín Fernández de Enciso, the navigator and geographer from Seville, placed the line of demarcation at a longitude corresponding to 45°38′ W in modern units, although his descriptions are less than clear. Harrisse (1897: 103–108, 122, 192–200) concluded that the line identified by de Enciso could also be near the mouth of the Amazon river between 49° and 50° W.

- Finally, two opinions were published in 1524. The Castilian captains Thomas Duran, Sebastian Cabot and Juan Vespuccius offered their insights to the *Badajoz Junta*, the authority charged with settling territorial disputes. However, the *Badajoz Junta* did not manage to reach an unambiguous decision. The captains suggested that the line of demarcation was located at 22° plus nearly nine miles West of the centre of the westernmost Cape Verde island, Santo Antão. This corresponds to a modern longitude of 46°36′ W (Harisse, 1987: 138–139, 207–208). Independently, Portuguese representatives presented a globe to the *Badajoz Junta*, showing the line at 21°30′ West of Santo Antão, or 22°6′ 36″ to the West of the island in modern units (Harisse, 1987: 207–208).

This focus on a demarcation line in the Atlantic Ocean only changed with the Portuguese discovery of the rich Maluku or Moluccas Islands—the Spice Islands—in 1512, an archipelago in present-day eastern Indonesia. In 1518, Spain insisted that the *Treaty of Tordesillas* divided the earth into two equal hemispheres, which would hence favourably place the Moluccas in its own, western sphere of influence. To resolve this potential conflict between the maritime powers, the *Treaty of Vitoria* (19 February 1524) charged the *Badajoz Junta* to meet that year to seek a solution to the disagreement on the 'anti-meridian'. That approach to reconciliation failed (Blair, 1903).

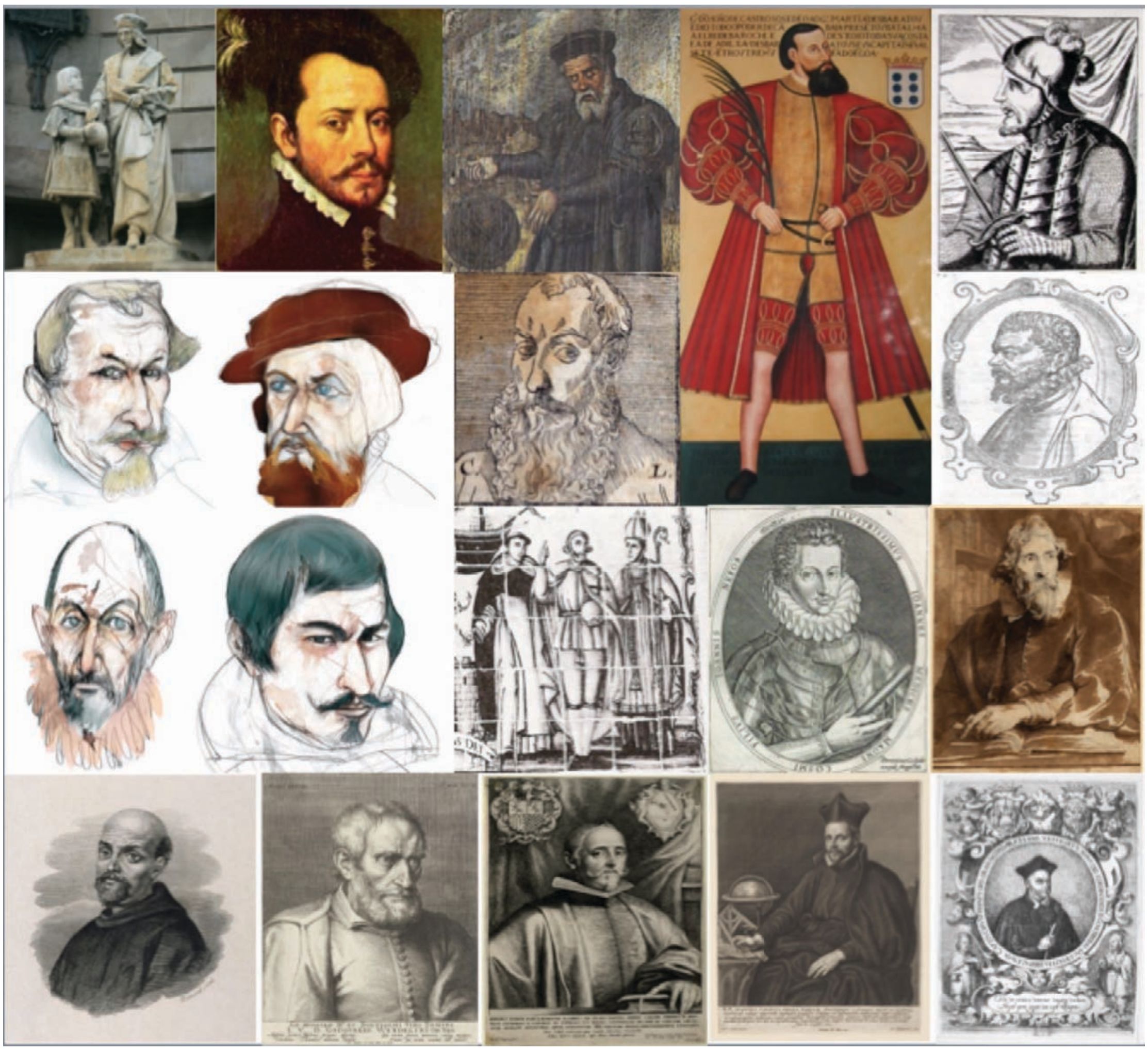

**Figure 6**: As Figure 1 but for the officials, cosmographers and service personnel playing important roles in the validation of all longitude solutions discussed in this article. *(i)* Ferrer de Blanes (1445–1523); *(ii)* Fernández de Enciso (1470–1528); *(iii)* de Medina (1493–1567); *(iv)* de Castro (1500–1548); *(v)* Nunes Salaciense (1502–1578); *(vi)* de Santa Cruz (1505–1567); *(vii)* Cortés de Albacar (1510–1582); *(viii)* de Chaves (1523–1574); *(ix)* Zamorano (1542–1620); *(x)* García de Céspedes (1560–1611); *(xi)* Cedillo Díaz (ca. 1565–1625); *(xii)* Ramírez de Arellano (1565–1633); *(xiii)* Giovanni de'Medici (1567–1621); *(xiv)* Puteanus (1574–1646); *(xv)* Castelli (1578–1643); *(xvi)* Wendelin (1580–1667); *(xvii)* Ramírez de Prado (1583–1658); *(xviii)* della Faille (1597–1652); *(xix)* de Contreras (1629–1685).
*Figure credits*: Wikimedia Commons, except for *(ii)*: https://alchetron.com/Mart%C3%ADn-Fernández-de-Enciso; *(iii)* Courtesy, Fundación Ignacio Larramendi; *(v)* https://www.pbslearningmedia.org/resource/bal111017eng/portrait-of-pedro-nunes-bal111017-eng; *(xviii)* National Gallery of Victoria, Melbourne, Australia/Everard Studley Miller Bequest, 1959. All images are in the public domain, except for *(ii)*, *(vi)*, *(vii)*, *(viii)*, *(x)*, *(xi)* and *(xv)*: Creative Commons Attribution-Share Alike 4.0 International license; *(ix)* Creative Commons CC0 1.0 Universal Public Domain Dedication.

Both parties would eventually reconcile, reaching an agreement that was enshrined in the *Treaty of Zaragoza*[7] (22 April 1529). Spain would retain its claim to the Philippines but forego taking possession of the Moluccas archipelago in return for a payment by Portugal of 350,000 ducats. In case Spain decided to return the money, the *Treaty* included a cancellation clause. The anti-meridian was set at 297.5 leagues or 17° to the East of the Moluccas. The *Treaty* also specified that the demarcation line passed through the islands of Las Velas (The Sails) and Santo Thome (Blair, 1903). The identity of the latter of these islands is unknown.

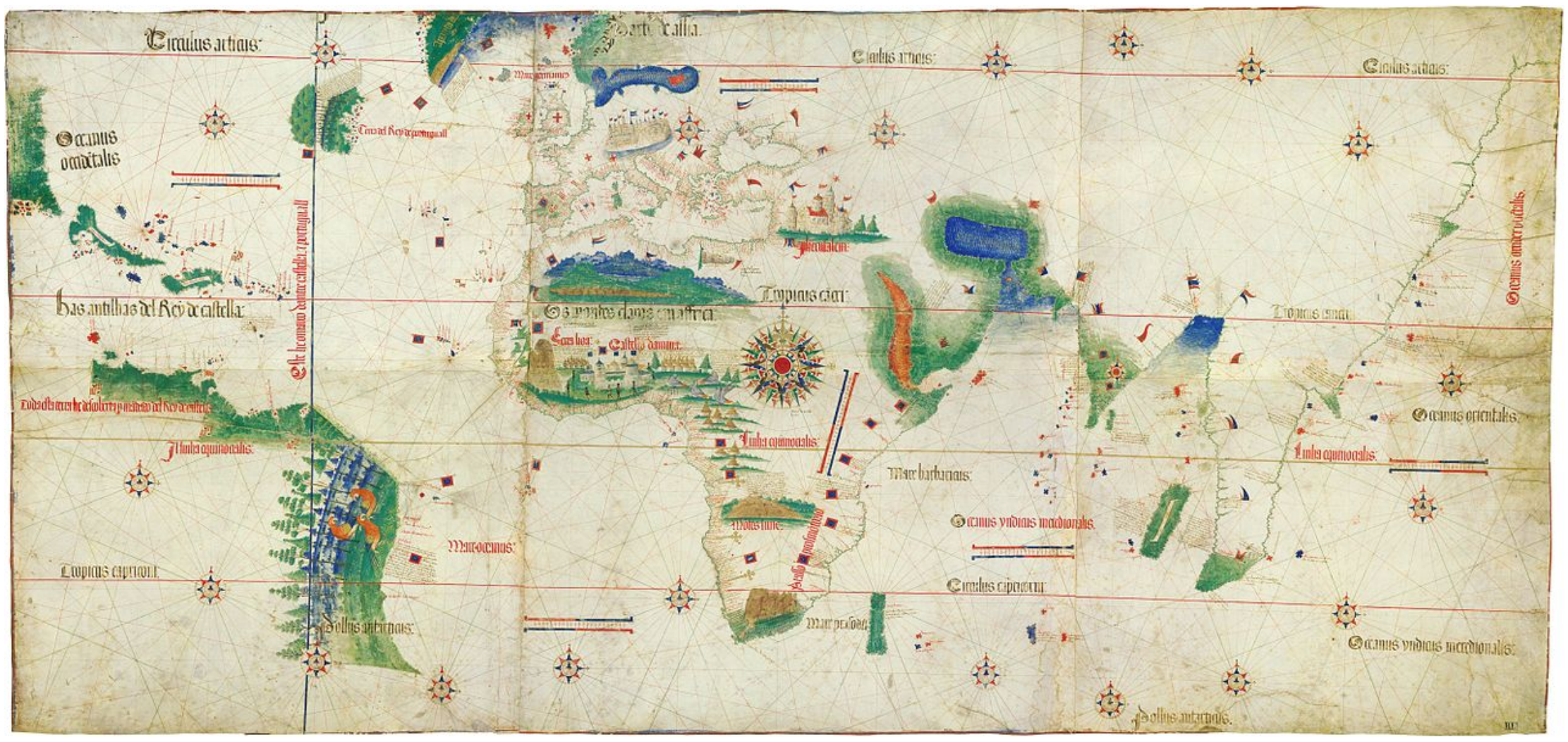

**Figure 7**: The *Cantino* planisphere, ca. 1502. (Biblioteca Universitaria Estense, Modena, Italy; public domain)

The Islas de las Velas, on the other hand, can be found in a Spanish history of China from 1585 (González de Mendoza, 1585), on the world maps of the Dutch cartographer Petrus Plancius (1594) and of Petro Kærio (1607), and in the 1598 London edition of Jan Huygen van Linschoten's collection of nautical maps. They correspond to a North–South island chain known in the sixteenth century as the Islas de los Ladrones, the Islands of the Thieves (Cortesao, 1939; Clark, 2005), later renamed the Mariana Islands. Guam, the southernmost and largest of the Mariana Islands, is located 17°21′ to the East of the Moluccas, thus confirming the islands' seventeenth-century identification. Figure 8 is a reproduction of an anonymous map from around 1550 showing the extent to which the coastlines of Eastern Africa, Asia and Western Oceania were known.

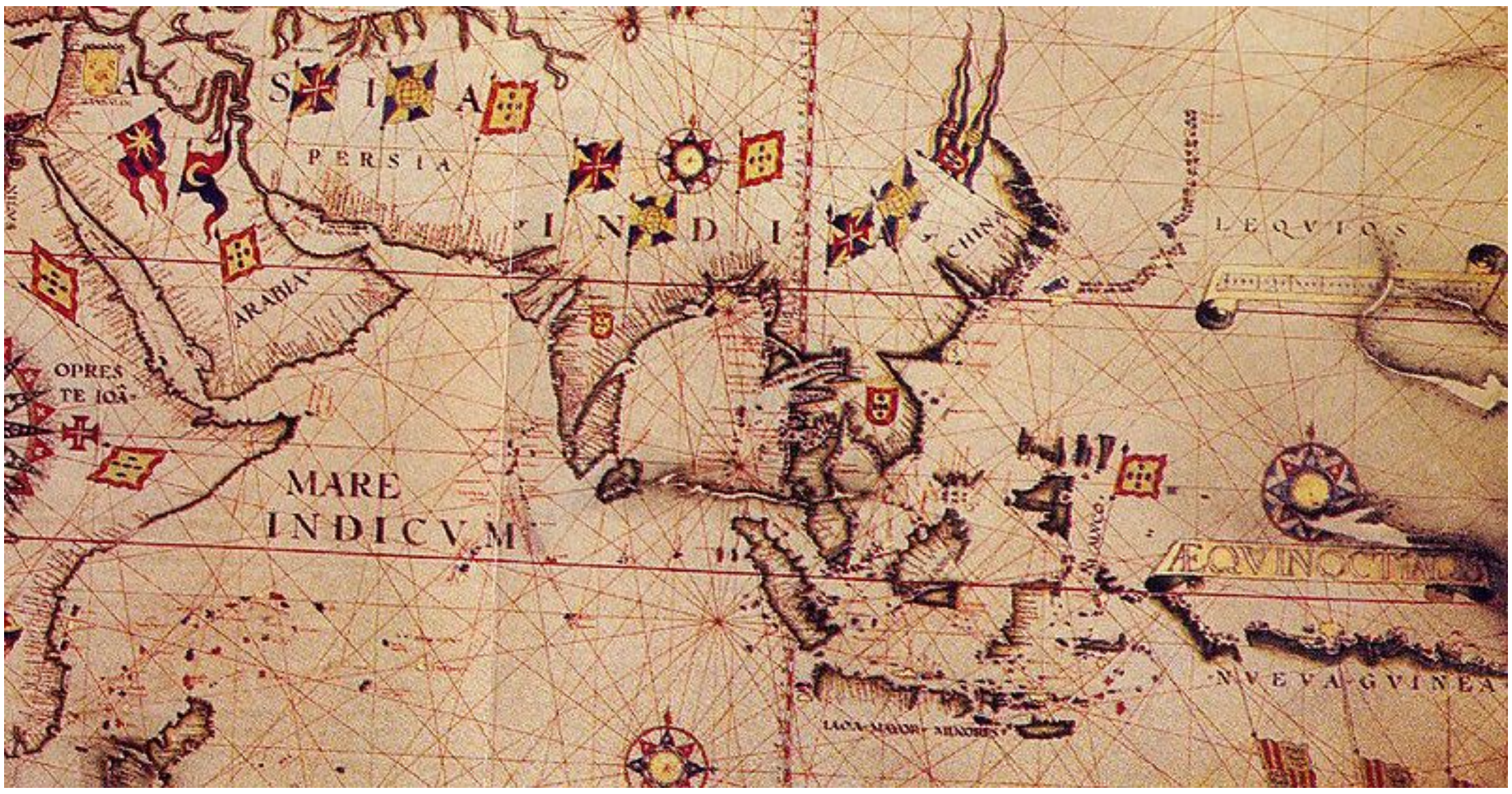

**Figure 8**: Anonymous map ca. 1550 of the known coastlines of Eastern Africa, Asia and Western Oceania. (Wikimedia Commons; public domain)

## 3 THE *CASA DE LA CONTRATACIÓN* AND THE *CONSEJO DE INDIAS*

In view of the prevailing disagreements about the precise location of the line of demarcation, the Spanish Crown was keen to pursue a policy of encouraging the independent development of accurate methods of longitude determination at sea (Fernández de Navarrete, 1802).

Already in 1567, King Felipe (Philip) II offered a reward of 6,000 ducats to anyone able to propose a practical and reliable solution to this thorny problem (Morato-Moreno, 2016).

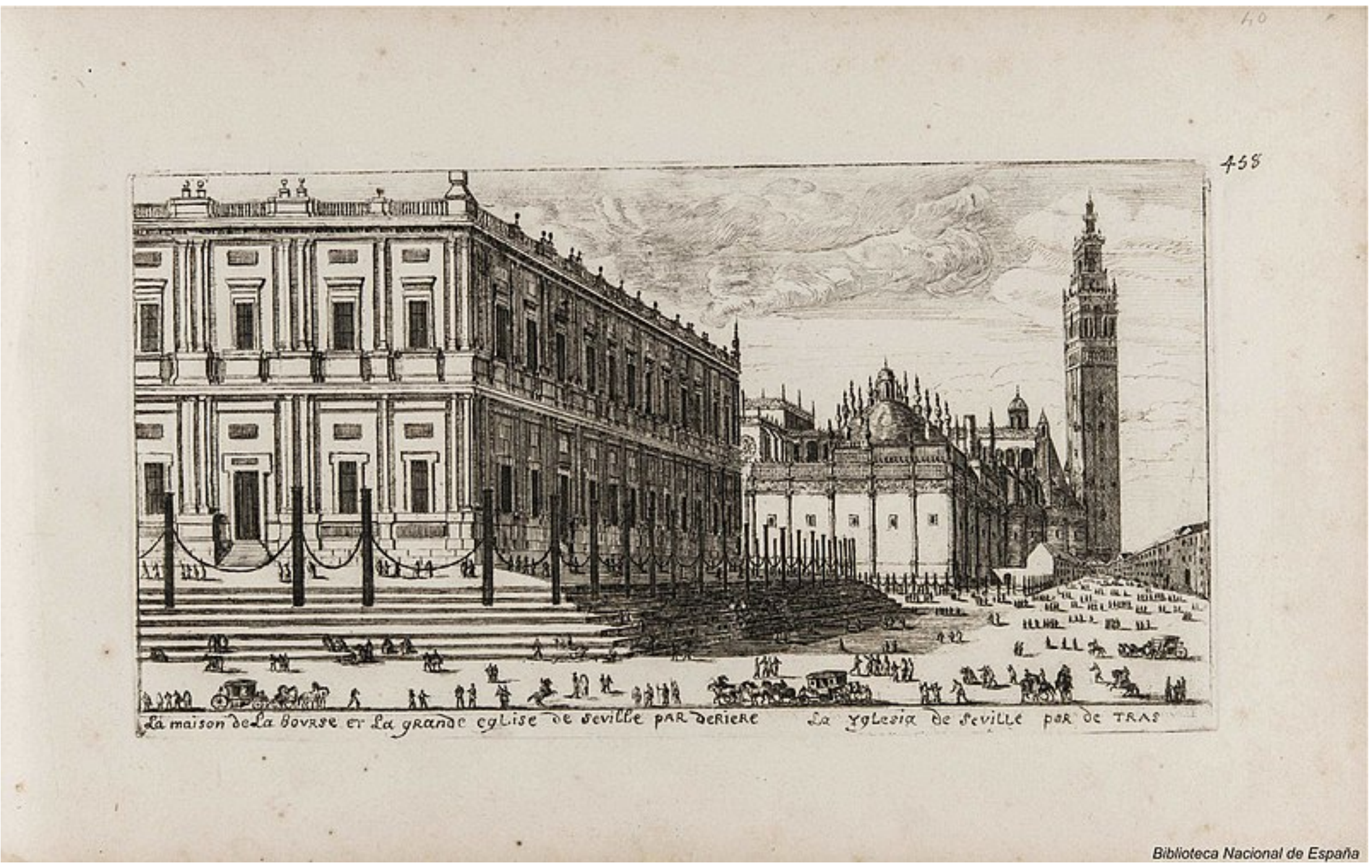

**Figure 9**: *Casa de la Contratación*, Seville. (Biblioteca Nacional de España via Wikimedia Commons; public domain)

Following his accession to the Spanish throne in 1598, King Philip III of Spain, Portugal and the Two Sicilies (that is, Sicily and Naples) followed suit. He was advised to offer a substantial prize to "the discoverer of longitude", thus testifying to the importance of solving the longitude problem. This prompted him to increase the prize money to 6,000 ducats in cash, a life annuity of 2,000 ducats and an expenses component of up to 1,000 ducats (O'Connor and Robertson, 1997). In today's money, one ducat was worth approximately €58 (Anonymous, 2014), so 6,000 ducats is equivalent to almost €350,000 today (Rodrigo, 2016).

The king was represented by the *Casa de la Contratación de las Indias* (see Figure 9)—the House of Trade of the Indies, formally *La Casa y Audiencia de Indias*—in Seville, a crown agency of the Spanish Empire. The *Casa de la Contratación* was entrusted with judging and validating any navigational proposals put forward. Ferdinand and Isabella had established the *Casa de la Contratación* in 1503 to regulate Spain's shipping and trade with the new Spanish American territories. From 1508 it became increasingly important as a centre of scientific and technical navigational expertise. A large number of mapmakers, or 'cosmographers', employed by the Spanish monarch supervised the development of nautical instruments, the preparation of nautical charts and the maintenance of a standard chart of the Indies. The latter, the secret *Padrón Real* (see Figure 10), was continually updated as new information surfaced (Goodman, 1991). Amerigo Vespucci, famous for his voyages to the New World, worked for the *Casa de la Contratación* as a pilot until he passed away in 1512; he was made *piloto mayor* (chief navigator) in 1508.

Perhaps the most important of these early royal cosmographers was Alonso de Santa Cruz, a celebrated mapmaker serving on the *Conseja de Indias* (Council of the Indies) and appointed to the *Casa de la Contratación* in the mid-1530s. In the late 1530s, de Santa Cruz presented the first map showing isogones—magnetic declinations, or variations from true North—insights he hoped to use to determine longitude away from well-known landmarks. His ill-fated pursuit of a magnetic solution to the longitude problem was presumably triggered by a suggestion from the Seville apothecary Felipe Guillén in 1525 (Ricart y Giralt, 1904; Sala Catalá, 1992). De Santa Cruz proceeded to develop a navigational instrument that used the effect of magnetic declination on the compass needle to determine longitude, which he presented to the *Casa de la Contratación* in 1536 (Cuesta Domingo, 2004; Portuondo, 2009). However, neither de Santa Cruz nor his contemporaries—most importantly, João de Lisboa, Pedro Nunes Salaciense and Guillén—ever managed to make it work. In fact, when his proposed solution failed to live up to actual measurements, obtained during the first geomagnetic survey compiled en route to the Indies by João de Castro (1539–1542), he conceded that “the whole idea of thinking that the longitude might be learnt … by means of the variation that the sailing-compass made, or that it produces them proportionally, left me” (de Santa Cruz, 1555). Nevertheless, a possible magnetic solution to the longitude problem caught the imagination of many would-be cosmographers, as we will see shortly.

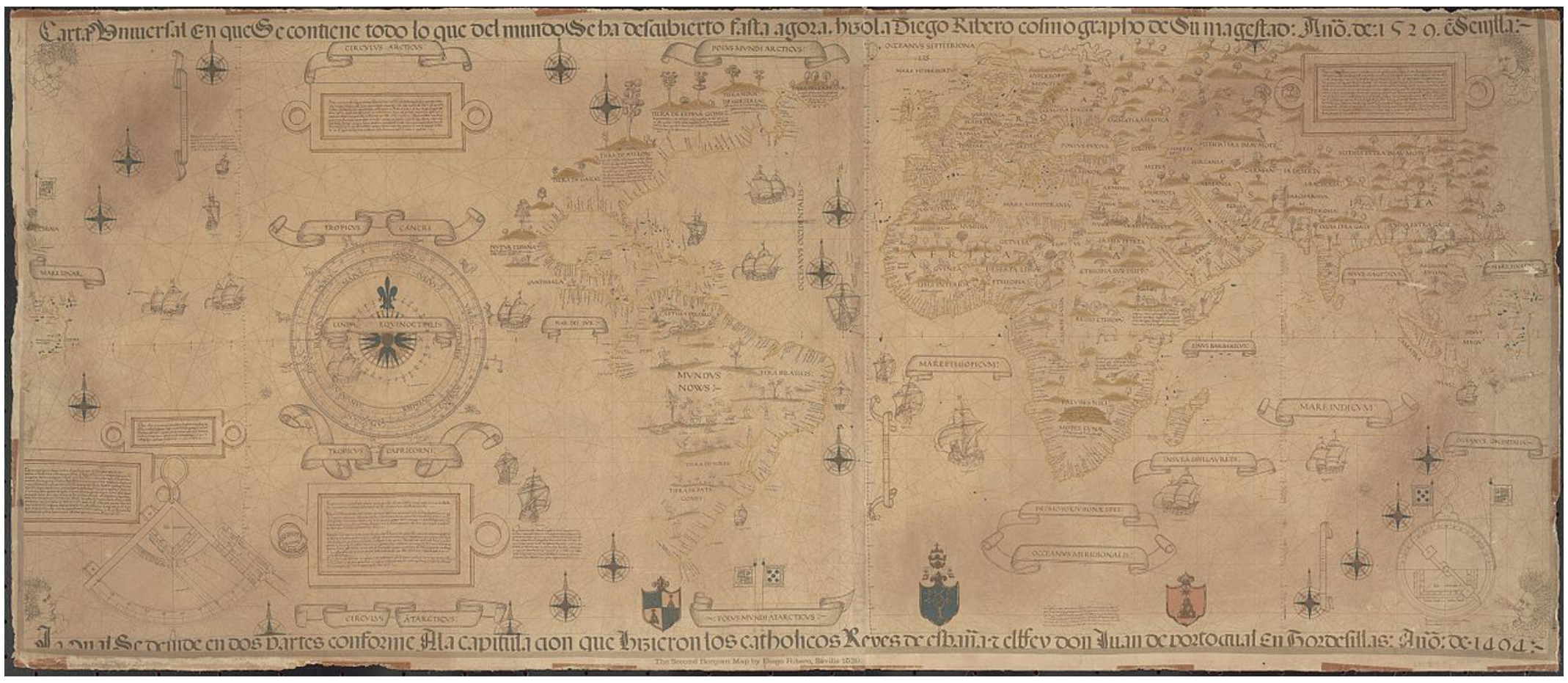

**Figure 10**: World map by Diego Ribero (1529); copy of the *Padrón Real*. (National Library of Australia via Wikimedia Commons; public domain)

The idea to base longitude determination at sea on the behaviour of the Earth’s magnetic field is often said to have originated from a discovery Columbus made on his transatlantic voyages.[8] Columbus had noticed that the direction of the compass needle (its ‘declination’) varied systematically on his voyage to the West. Upon his return to the Iberian Peninsula, this suggestion led to a heated debate among contemporary scholars, essentially dividing the experts into two opposing camps. High-profile cosmographers, including Martín Fernández de Enciso, Nunes and Pedro de Medina vehemently opposed Columbus’ suggestion, but others such as Francisco Fale(i)ro and Martín Cortés de Albacar began large-scale efforts to tabulate and compile compass measurements as a function of location. This eventually led to Cortés’ formulation of the concept of an Earth ‘magnetic pole’, which became compulsory reading for students of terrestrial magnetism in the sixteenth century:

> Many and diverse are the opinions that I have heard, and from some modern writers I have read about the northeastern and northwestern [pointings] of the needles; and it seems to me that none work properly and few are correct. They say northeast when the needle points north to northeast, and northwest when it tends to the northwest. To understand such differences in which needles differ from the [direction to the] pole, imagine a point below the world’s pole and this point is outside all the skies that are under the prime mover.[9] Which point or part of the sky has an attractive property that thus attracts a magnetized iron [needle] ... this point is not located in the moving skies nor is it at the pole, because if it were, the needle would not point northeast- nor northwestwards. (Sala Catalá, 1992)

In 1552, King Philip II created an academic ‘chair of the art of navigation and cosmography’ at the *Casa de la Contratación*, with Jerónimo de Chaves (Chavez) as first professorial office holder. His main task was the formal instruction, examination and licensing of maritime masters and pilots on the route to the Indies. The emphasis of his instruction was placed on setting a course on the basis of marine charts, determination of latitude from observations of the Sun and Polaris, the use of clocks, understanding tides and the theory and practice of nautical instruments “so that errors in them can be detected” (Portuondo, 2009). The *Casa*’s teaching manuals, particularly the *Arte de Navegar* (Valladolid, 1545; see Figure 11) of Pedro de Medina, were translated widely. Figure 12 is a representation of the vessels of the early Spanish navigators found in de Medina’s *Arte de Navegar*, re-engraved in the Venice edition of 1555.

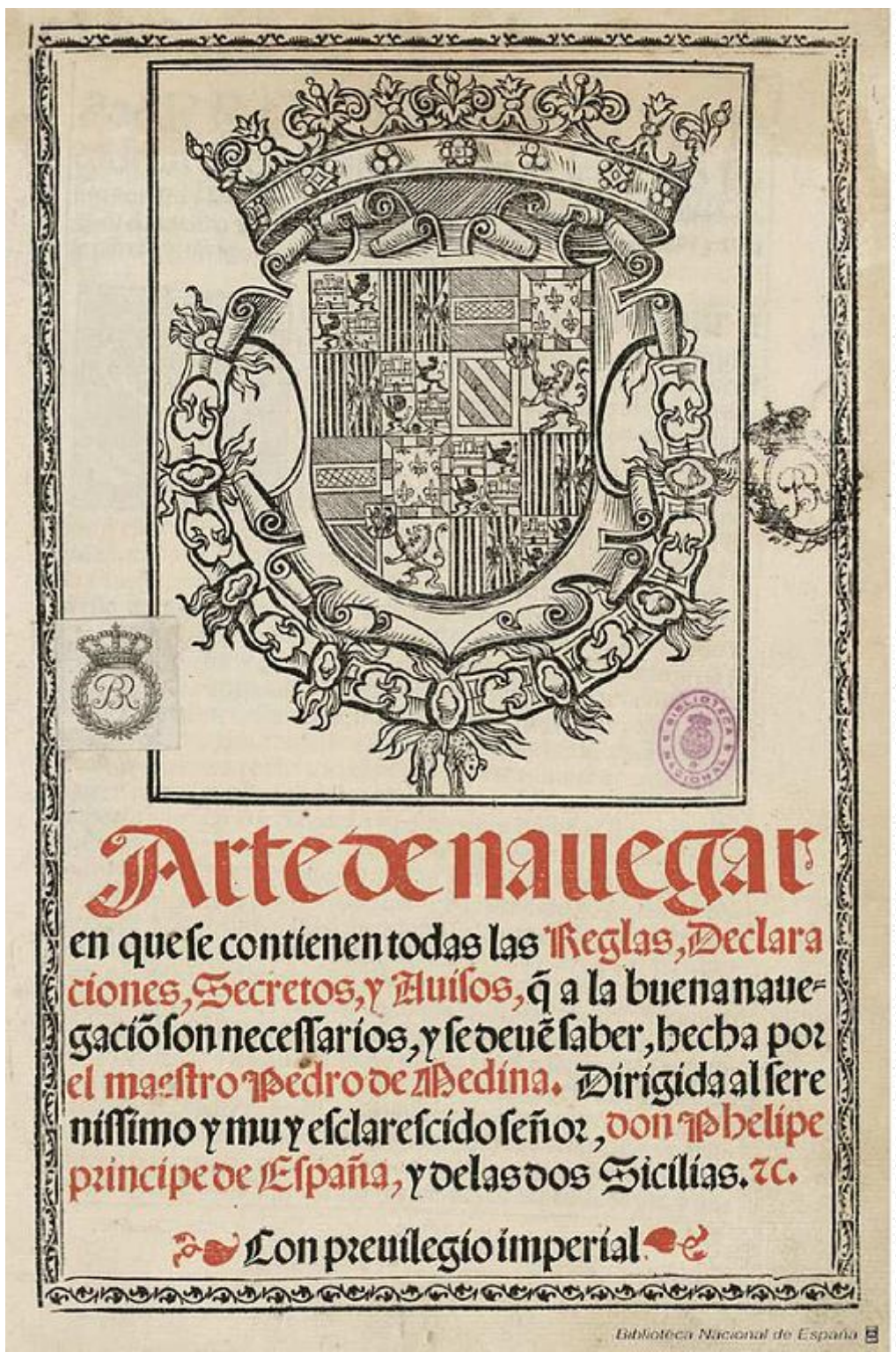

**Figure 11**: *Arte de Navegar* (Valladolid, 1545) by Pedro de Medina. (Biblioteca Nacional de España; Creative Commons Attribution-Share Alike 4.0 International license)

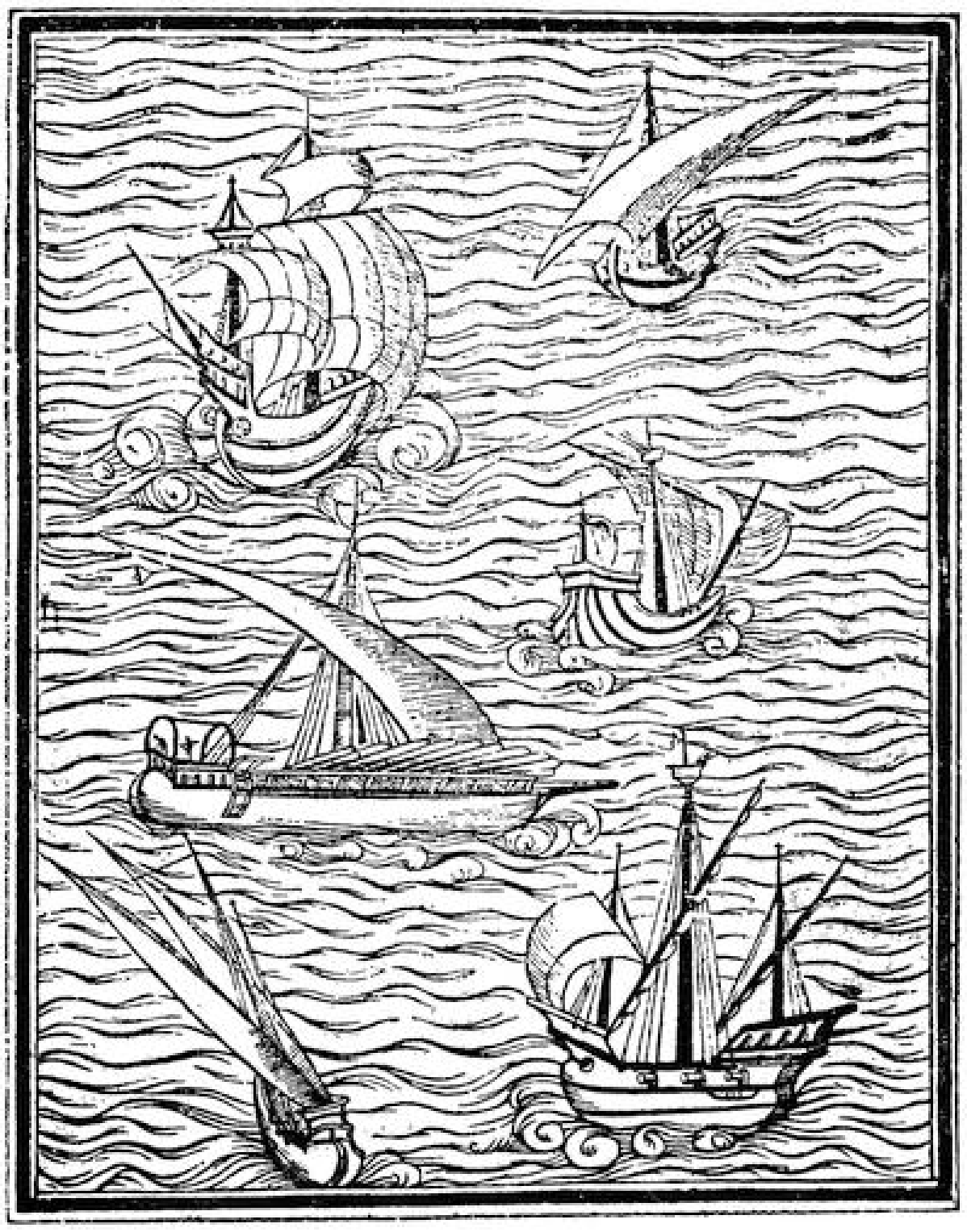

**Figure 12**: Representation of the vessels of the early Spanish navigators, from de Medina's *Arte de Navegar*, re-engraved in the Venice edition of 1555. (Winsor, 1886; public domain)

Meanwhile, and given his favourable standing on the *Conseja de Indias*, de Santa Cruz was called upon in 1554 to assess a proposal for a device to determine longitudes submitted by Petrus Apianus. He concluded that the proposed instrument was, in fact, rather similar to his own, earlier invention. Shortly afterwards, in 1555 de Santa Cruz published his *Libro de Longitudes* (Book of Longitudes; thought to be based on a preliminary version presented to the *Casa de la Contratación* as early as 1539). The treatise contained descriptions of 12 methods of longitude determination proposed since Antiquity, including the use of lunar and solar eclipses, which were promising provided that one could accurately determine one's local time based on astronomical observations and compare the measurements with known ephemerides:

> That the eclipses were well calculated by men trained in astrology before their departures to know precisely the day and hour and point of it, in which they had to start or end such eclipses, they could find out quite precisely the length of any place based on the point where they left.

However, given their rarity and because of the difficulties of determining the exact times of ingress and egress, he suggested that such methods might only feasibly be employed on land.[10] (For a detailed discussion of the use of lunar eclipses as a contemporary means of longitude determination, see de Grijs, 2020.) De Santa Cruz concluded, well ahead of his time, that longitude determination at sea would only be feasible if one had access to an accurate timekeeper. After all, the Earth performs one full rotation, 360 degrees, in 24 hours, so each hour it traverses 15 degrees. Thus, if one could accurately determine both one's local time at sea (e.g., using astronomical observations or solar altitude measurements) and the local time at a predetermined reference position, the time difference would directly correspond to the difference in longitude between both locales. This was not an altogether novel suggestion, however. In 1524, Ferdinand Columbus (Hernando Colón), son of Christopher Columbus, had already emphasised the need to equip the Admiral's fleet with accurate clocks to calculate time differences with respect to a reference meridian.

To date, historians continue to debate the novelty of de Santa Cruz's *Book of Longitudes*. Assessments range from it representing an important benchmark in the history of navigation to blatant plagiarism (Cuesta Domingo, 2004). One example of the inaccuracies in de Santa Cruz's account is provided by his estimate of the difference in longitude between Mexico City and Genoa, Italy, of 217°30', of order twice the actual difference in longitude (Goodman, 1988: 55).

**4 EARLY CONTENDERS**

Following King Philip II's announcement of his longitude prize, a flurry of activity ensued (Barrado Navascués, 2014). The situation in late-sixteenth-century Spain is reminiscent of that in early-eighteenth-century Britain, with numerous half-baked proposals seeing the light in pursuit of the country's coveted longitude prize (e.g., Andrewes, 1996; de Grijs and Vuillermin, 2019). This resulted in significant expenses as well as abuse, thus causing the Spanish kings to distrust any promises from those who claimed to have found a solution. The celebrated Spanish author, Miguel de Cervantes, and some of his contemporaries began to ridicule these attempts to find *el punto fijo* (the 'fixed point', i.e., one's longitude) at sea (e.g., de Cervantes, 1615). Similarly as in Britain, 'finding the longitude' took on the meaning of striving to accomplish the impossible, just like 'squaring the circle'. As a case in point, consider the following passage attributed to de Cervantes' mathematician in *El coloquio de los perros* (*The Dialogue of the Dogs*, 1613):

> I have spent twenty-two years searching for the fixed point [*el punto fijo*] and here it leaves me, and there I have it, and when it seems I really have it and it cannot possibly escape me, then, when I am not looking, I find myself so far away again that I am astonished. The same thing happens with squaring the circle, where I have arrived so near to the point of discovering it that I do not know and cannot imagine how it is that I have not got it in my pocket... (de Cervantes, 1944)

By 1570, one Juan Alonso from Gran Canaria had constructed a navigational instrument resembling an astrolabe,

> … for measuring the height of the Sun at any time of day …, for knowing the distance between places according to [their] longitude[s] without observing eclipses … and for navigating East to West with remarkable ease and certainty. (Fernández de Navarette, 1852)

Alonso had spent most of his adult life studying mathematics, and he had been forced to overcome many difficulties to eventually construct his novel instrument. He had hoped to demonstrate his invention to the king in person, but he was prevented from doing so because of a serious and paralysing illness he contracted in 1566. As luck would have it, however, Dr. Fernán Pérez de Grado, Regent of the *Audiencia* in the Canary Islands, was persuaded as to the device's promise and hence Pérez de Grado notified the king.

King Philip II subsequently issued a royal decree, dated 4 August 1571, for Alonso to organise transportation of his instrument and any relevant documentation to the Spanish court in Madrid by a trusted person. The king saw fit to promise Alonso prize money, by proxy through Pérez de Grado, if his invention's performance held up to careful scrutiny by the *Consejo de Indias* during sea trials. Seven months later, Pérez de Grado's two sons—Alvaro and Alonso—travelled to the royal court. The Regent's sons took along with them Alonso's written *método* and a letter dated 15 March 1572. In his written missive, Alonso implored the king to have a number of additional devices constructed in case something went wrong with his single existing instrument. He also took the opportunity to subtly request that his efforts be rewarded financially if appropriate, so as to alleviate his family's poverty.

Prior to the departure of the Regent's sons, Alonso had instructed the young men in the practical operation of his device. In his letter to the king, Alonso had specifically asked for "honest men of science" to examine the performance of his device, recommending his youngest son, Alonso, an experienced sailor, for the instrument's impending sea trials. Unfortunately, there are no contemporary sources of information regarding the device's fate, or that of the sea trials, if any. In the words of Fernández de Navarette (1852), "We have been unable to find out more about the success of this invention, nor about any experiences that should have qualified it" (but for a speculative link, see Fernández de Navarette, 1852: 12). Yet, if Alonso's invention and method were deemed useful by the cosmographers of the *Consejo de Indias*, this episode could well have been an important next step in the verification and development of operational instrument designs by the likes of Juan de Herrera, for instance (Fernández de Navarette, 1852: 13, 14; Marquez, 2017).

De Herrera, the well-known architect of El Escorial, the residence of the Spanish monarch, was given the *privilege* of manufacturing an instrument for longitude determination by King Philip II on 13 December 1573 (Barrado Navascués, 2014). His star rose rapidly, since the following year it was decreed that the instrument would be provided to the Spanish India fleet. Yet, despite the apparent practical advantages the device afforded, de Herrera did not receive any share of the Spanish longitude prize. He embarked on an illustrious career, however, since in 1582 the Spanish Academy of Mathematics was established under his direction, with cosmography featuring as one of its main areas of interest.

Contrary to his father's attitude, King Philip III took his responsibilities as king less seriously. Moreover, he was not particularly impressed by any of the schemes proposed (O'Connor and Robertson, 1997). Nevertheless, the cosmographers of the *Casa de la Contratación* considered each proposal on its merits. Of the ten credible proposals submitted by 1634, two were fronted by scholars not already associated with the core of the Spanish empire (Fernández de Navarette, 1852; Randles, 1985). I will discuss these, as well as the eight other proposals submitted to the *Conseja de Indias* in appropriate detail below.

## 5 IBERIAN SQUABBLES

### 5.1 Magnetic declinations

The first decade of the seventeenth century saw a flaring up of tensions between two prominent proponents of the magnetic solution to the longitude problem. The dispute at the heart of the tensions, a technical altercation between Juan Luis Arias de Loyola on the one hand and the Portuguese hopeful Luís da Fonseca Coutinho on the other, played out before the *Conseja de India* and, eventually, led to a significant waste of resources, personal squabbles and scientific bitterness (Baldwin, 1980).

Formerly a professor of mathematics at the Academy of Mathematics, Arias de Loyola was appointed in 1591 to the *Conseja de Indias* as chronicler and cosmographer, succeeding Juan López de Velasco. This role had been created to predict and observe lunar eclipses and for compiling the general and natural histories of the Indies, among other tasks. However, in 1594 the dual roles of chief chronicler and chief cosmographer were recast into two new positions. Soon afterwards, by 1595, Arias de Loyola was relieved of his functions on account of the poor quality of his work as chronicler. He was replaced by Pedro Ambrosio de Ondériz. Yet, despite his demotion, he continued his career as cosmographer at the court in Valladolid, where he was well placed with access to the Spanish monarch.

In 1603, Arias de Loyola announced that he had discovered five "secrets of navigation", among which the most promising and far-reaching was "determination of the longitude, which sailors commonly call navigation from East to West" (Fernández de Navarette, 1852: 16; Vicente Maroto, 2001). As the importance of resolving the longitude problem could not be overstated, the king felt obliged to solicit a technical assessment of this claim from the *Conseja de Indias*. The *Conseja* summoned Arias de Loyola on 24 July 1603 to disclose the technical details of his newly proposed method so that skilled mathematicians and experienced pilots could assess its promise. However, Arias de Loyola was loath to reveal his secret if he was not guaranteed a significant cut of the prize money. The *Conseja* agreed that if the method turned out to be useful and practical, Arias de Loyola would be rewarded accordingly. His method was slated for assessment by the cosmographer and chief pilot of the *Casa de la Contratación*, Rodrigo Zamorano, two additional mathematicians from Seville, as well as a number of generals and other pilots. When the king received their report on 2 September 1603, he ordered that Arias de Loyola be given 600 ducats for expenses, and that João Bautista (Baptista) Lavanha and Andrés García de Céspedes, cosmographers at the Spanish court, proceed to evaluate the proposal in detail (Fernández de Navarette, 1852: 16).

Although Arias de Loyola's proposal was presented to the *Casa de la Contratación* already in 1603, Fonseca Coutinho's submission, presented to the king some time in 1604 or 1605 (Fernández de Navarette, 1852: 17; Vicente Maroto, 2001), was the first to be offered a significant cut of the Spanish longitude prize. The *Conseja de Indias* formally declared its intention to proceed with Fonseca Coutinho's proposal on 11 August 1607 (Fernández de Navarette, 1852: 17). The instruments proposed by both men were largely similar, and in fact Arias de Loyola's design was more precise. Yet, Fonseca Coutinho had the ear of the chief cosmographer, Lavanha, and his proposal, supported by the Crown of Portugal, was hence first brought before the king (Ceballos-Escalera Gila, 1999). Arias de Loyola's representation was delayed time and again on various pretexts. Fonseca Coutinho's proposed verification of his method required the *Casa de la Contratación* to purchase 20 to 30 compasses in both Seville and Lisbon for use on the India voyages. Fonseca Coutinho proposed to equip each ship with six compasses, three ordinary and three box compasses, so as to be able to properly compare the measured compass needle variations during the voyage. This may well have been one of the first applications of statistical analysis in the history of science.

However, it was not Fonseca Coutinho himself but Blaz Telles, Admiral of the Spanish India fleet, who was tasked with the acquisition of the devices, at great cost: the Lisbon order set the Portuguese Crown back 300 ducats. Six new astrolabes were added to the instrumentation list, at a cost of 40 ducats each, to which cartographic aids worth 100 ducats were also added. Fonseca Coutinho became highly adept at finding pretexts and excuses as to why he could not undertake the voyage and acquire the instruments himself, variously citing practical and financial difficulties (Fernández de Navarette, 1852: 17). Unfortunately, Telles reported that all compasses had yielded 'fixed' results on both legs of the voyage, a similar outcome to that resulting from the use of compasses examined by Fernando de los Ríos, Attorney General of the Philippines. Fonseca Coutinho was afforded the chance to provide redesigned compass needles for a second trial, which was to commence in 1609 (Ceballos-Escalera Gila, 1999). His new design included both 'regular' and 'equinoctial' needles. The former would indicate any deviation to the East or West with respect to the reference meridian, while the latter was "placed in a graduated ring, and supported on two axes, kept from the air by two glasses, the instrument freely hanging" (Ceballos-Escalera Gila, 1999).

The *Conseja de Indias* suggested to the king that Fonseca Coutinho be offered a generous award (Baldwin, 1980): 2,000 ducats from the *avería*—a tax on the Indies trade that covered the costs of providing armed protection for merchant shipping—as well as 2,000 ducats from the *real hacienda* (the royal treasury) and 2,000 from the Crown of Portugal, combined a total monetary value of 6,000 ducats. King Philip III agreed, although he proposed that the 2,000 ducats that were to be paid by the royal treasury should be divided between the *Casa de la Contratación* and the Crown of Portugal, given the enormous pressures the royal treasury was facing already. In addition, he offered a number of personal presents and honours, plus 1,000 ducats to cover expenses. In response, the *Casa* held firm on the royal treasury's contribution to the award, pointing out the king's significant interest in and potential benefits of resolving the longitude problem (Fernández de Navarette, 1852: 18). The king was also asked to appoint an official to whom Fonseca Coutinho would divulge his secrets, given the latter's advanced age and poor health. Despite these arrangements, the Portuguese projector[11] refused to disclose the details of his method, although around the time he first submitted his proposal to the *Casa de la Contratación*, he had compiled an unpublished manuscript (Ceballos-Escalera Gila, 1999; Turley and Souza, 2017: 131, note 229), *Arte de agulha fixa, e do modo de saber por ella a longitud* (The Art of the Fixed Needle and the Method of Finding the Longitude with It).[12]

Fonseca Coutinho's good fortune turned on 8 June 1610, however, when land-based and sea-borne technical assessments involving the best cosmographers and pilots—including Lavanha, Jerónimo de Ayanz y Beaumont ('Ayanz'), Juan Cedillo Díaz, Alonso Flores, Antonio Moreno, Remando de los Ríos and T(h)omé Cano (Vicente Maroto, 2001)—led to the conclusion that his approach was unsound (see also Ceballos-Escalera Gila, 1999: 28–34).

Instead, the *Conseja* eventually determined that Arias de Loyola should be awarded the longitude prize, in addition to the 600 ducats he had already been offered to cover expenses.

Arias de Loyola had finally secured the support of a powerful ally in the person of Pedro Fernández de Casto y Andrade, Great Count of Lemos and president of the *Consejo de Indias* (Barceló, 2019). As a consequence, Arias de Loyola was finally heard by the authorities. The cosmographer asked for a reward of 10,000 ducats. However, on the recommendation of Zamorano and Lavanha, on 13 July 1612 the king signed a royal decree granting Arias de Loyola, his heirs and successors a perpetual income of 6,000 ducats, in addition to a personal life annuity of 2,000 ducats (Barceló, 2019), meanwhile nullifying Fonseca Coutinho's earlier reward (Vicente Maroto, 2001). Despite his solid credentials, Fonseca Coutinho's reputation was irreversibly damaged. Until the present time, he is predominantly seen as an *arbitrista* (schemer) and *embaucador* (charlatan).

Despite this encouraging turn of events, Arias de Loyola was not satisfied and felt that he had been cheated out of the prize money he felt he deserved. He penned his *Tratado del modo de hallar la longitud y la aguja fija*[13] (*Treatise on Finding the Longitude and the Fixed Needle*) about the innovative compass he had originally presented for assessment in 1603. In addition, he suggested the use of a second novel method of longitude determination that clearly anticipated application of Galileo's new telescope to follow the ephemerides of Jupiter's moons. I will examine Galileo's contribution to the longitude solution in Section 6.

However, after new land-based experiments by Cedillo Díaz, successor to García de Céspedes, and Lavanha, the proposal of Arias de Loyola was found lacking in precision too. And so, in 1615, King Philip III issued a decree offering Lorenzo Ferrer Maldonado, now commonly regarded a charlatan and trickster, 5,000 ducats for his secret of the …

> … fixed needle at all meridians of the world and the *punto fijo* of the longitude of navigation from East to West at any time of day and night, without sun or stars, and the secret of navigation from East to West. (Vicente Maroto, 2001)

Ferrer Maldonado's reward was pending practical verification, full disclosure of all technical details and the development of a useful means of practical implementation. He had lost much credit and earned an infamous reputation at the Spanish court given his propensity to propose the development of numerous fanciful devices that came to nothing. One of these, proposed with the aim to make everyone believe that he was an experienced and brilliant navigator and scholar, was his 'invention' of the first fixed compass that worked such that one could always find one's local longitude on Earth.

In compliance with the king's request for verification, Labanha and his fellow cosmographers at the *Conseja de Indias*, Francisco Garnica, Lúcas Guillén de Vea and Cedillo Díaz, proposed that Ferrer Maldonado embark on a voyage accompanied by two Portuguese pilots and two Castilians from Cádiz via the coast of Africa and the Cape of Good Hope to Santa Elena, then crossing the South Atlantic Ocean to Buenos Aires, following the coastline up to Brazil until Cuba, Santo Domingo and other Caribbean islands, eventually returning via the Azores to their home port in Spain (Fernández de Navarette, 1852).

Shortly after Ferrer Maldonado's promising presentation to the king, in September 1615, the French captain Juan de Mayllard also offered to divulge his secret of determining the longitude at any time of the day during sunshine hours. In response, King Philip III ordered that de Mayllard join Ferrer Maldonado on his voyage circumnavigating the South Atlantic, in order to compare the performance of both of their methods (Vicente Maroto, 2001). Ferrer Maldonado agreed, despite his inability to deliver. He asked for an advance, as did de Mayllard, so as to allow him to prepare for the voyage and construct the precise instrumentation he needed for his sea trials, all the while finding pretext after excuse to stall. The voyage never materialised. Compared with some of his other outrageous claims, however, including the suggestion that he had found the 'key of Solomon', having worked out how to convert any metal into gold, his navigational fakery seems almost benign.

Although Arias de Loyola had expressly requested that his technical innovations be treated confidentially, particularly with respect to the skillful Flemish competition (including

Michael Florent van Langren; see Section 7), he did not stick to his guns. In 1633, while complaining about the treatment he had received from the *Conseja*, he also revealed that he had been offered 100,000 escudos in gold to disclose those closely held technical details to a foreigner (Baldwin, 1980).

The initial dismissal of his innovations by the *Conseja de Indias* clearly continued to bother him for a long time. While Fonseca Coutinho suddenly disappeared from Madrid around 1610, Arias de Loyola stayed for the better part of another three decades, airing his grievances, issuing memoranda and attempting to discredit anyone who dared to contend the Spanish longitude prize (Roscoe, 1839: 181–182; Vicente Maroto, 2001). Unwitting victims of his spite included one Benito Escoto from Genoa in 1616, the Flemish mathematician and cosmographer at the Spanish court, van Langren, in 1632 and José Moura Lobo, the Portuguese explorer and cosmographer, in 1637.

Despite the early lack of progress in employing the compass needle in securing a resolution to the longitude problem, the *Casa de la Contratación* continued to sponsor voyages aimed at trialing the magnetic concept for a number of years. Cedillo Díaz had reported on the unsuccessful outcomes of the proposals submitted by Arias de Loyola, Fonseca Coutinho and Ferrer Maldonado. Therefore, he was the right conduit for assessment of similar methods trialed on a subsequent voyage by the brothers Bartolomé and Gonzalo García Nodal, as well as Diego Ramírez de Arellano, pilot and senior cosmographer of the *Casa*, upon his return from their expedition to the Straits of Magellan and San Vicente in 1618–1619 (Vicente Maroto, 2001). Ramírez de Arellano left a handwritten navigation log, which greatly facilitated Cedillo Díaz's assessment. None of these later voyages resulted in a viable means of longitude determination on the basis of magnetic declination, however.

### 5.2 Ayanz' mechanical approach

While Arias de Loyola and Fonseca Coutinho were hatching their plans to find a solution to the longitude problem by means of magnetic declination, Ayanz took a radically different approach. Lauded as a practical engineering genius who surpassed even Leonardo da Vinci (García Tapia and Cárdaba Olmos, 2013), Ayanz pushed a rather curious yet ultimately unworkable invention to determine the distance travelled at sea. He suggested to attach external wheels that revolved as the vessel made its way towards its destination. During the voyage, the number of revolutions made by the wheels should be counted and converted into the distance covered:

> Make a wheel with its planks at the front, about one *vara* [approximately 84 cm] in diameter or less, positioned between two supports and with forty-eight cogs, and this large cogwheel in a spindle that has six cogs in the small cogwheel; the wheel has to be set on the side where the rudder is, in such a way that once the vessel is loaded it does not sink too deep into the water and the spindle is what engages and raises the amount that is wanted, where, through a small gate, it engages another cogwheel, which has to engage the aforementioned 6-bar spindle; engages another wheel with forty-eight, in such a way that this is reached by multiplying that, when the wheel that is one *vara* in diameter revolves once, the ship travels three *varas* and when the second wheel turns once it travels the same distance, because the spindle that revolved eight times, makes the wheel that is above in the small gate rotate one more time and that wheel, beating against the other with the six cogs cogwheel, rotates eight times while the one below only turns once, and further wheels can be applied, so that the vessel travels ten thousand *varas* first with one turn of the wheel; and each one controls the number like a clock, knowing the number of times each one has revolved. And there is another similar wheel for the currents acting crossways at the stern, in such a way that any part of the of the ship's side that picks up the current, turns the wheel. Take note that the wheels that thrash in the water have to be covered at the sides, and the same applies to all that is above the waterline, so that the waves cannot beat against them, and the current created by the ship cannot beat against the wheel that is located crossways for the currents. And all that has to be done to find out how far the vessel has travelled is to count the number of times the main wheel has turned, and then look how much the other has been diverted by the current, and make two lines from the point, one straight line for the route, and the other for how much it has been diverted, and plot a circle from the point and measure the distance from the straight line to the diversion, and you will find the amount. And take care to look at the number of times the wheel revolves with a favourable wind and, if the wind changes direction and makes the vessel modify its route, write down the time that this happens, and how long this lasts, and count the times that the wheel rotates and the number of turns will tell you how far the vessel has been diverted.

In Valladolid on the twelfth day of March, in the year one thousand six hundred and two, Jerónimo de Ayanz. (García Tapia and Cárdaba Olmos, 2013: 93–94)

Ayanz was awarded a *Cédula de Privilegio* (a privilege of invention) by King Philip III on 16 June 1603, that is, the king offered him the privilege to enjoy the exclusive right of up to 50 inventions. Forty-eight inventions were recorded in his *Discurso* of 1 September 1606, which is preserved in both the Archives of Simancas[14] and the *Archivo General de Indias* in Seville (García Tapia and Cárdaba Olmos, 2013; García Tapia, 2016). *Cédula* documents are equivalent to authentic patents of invention, established following similar procedures of assessment as those in use today around the world.

Although the majority of his engineering inventions were indeed endorsed by the relevant authorities and signed off by the king, his solution to the longitude problem raised serious questions:

The next device cannot be adjusted because the sea currents that are at the rear of the vessel are lighter and because the wheel is lighter than the vessel, so it is not possible to know exactly how far the ship has travelled; and the same drawback applies, albeit to a lesser extent, to the ones that are arranged from stern to bow, although it is possible to know a bit more accurately how far the vessel has travelled and how far it has been diverted, and its approximate whereabouts; and as far as the currents are concerned, with experience it is possible to see how many more times the wheel has turned than the distance the ship has really travelled. If the ship had been diverted two leagues and the wheel had registered four, subtract the two. (García Tapia and Cárdaba Olmos, 2013: 93)

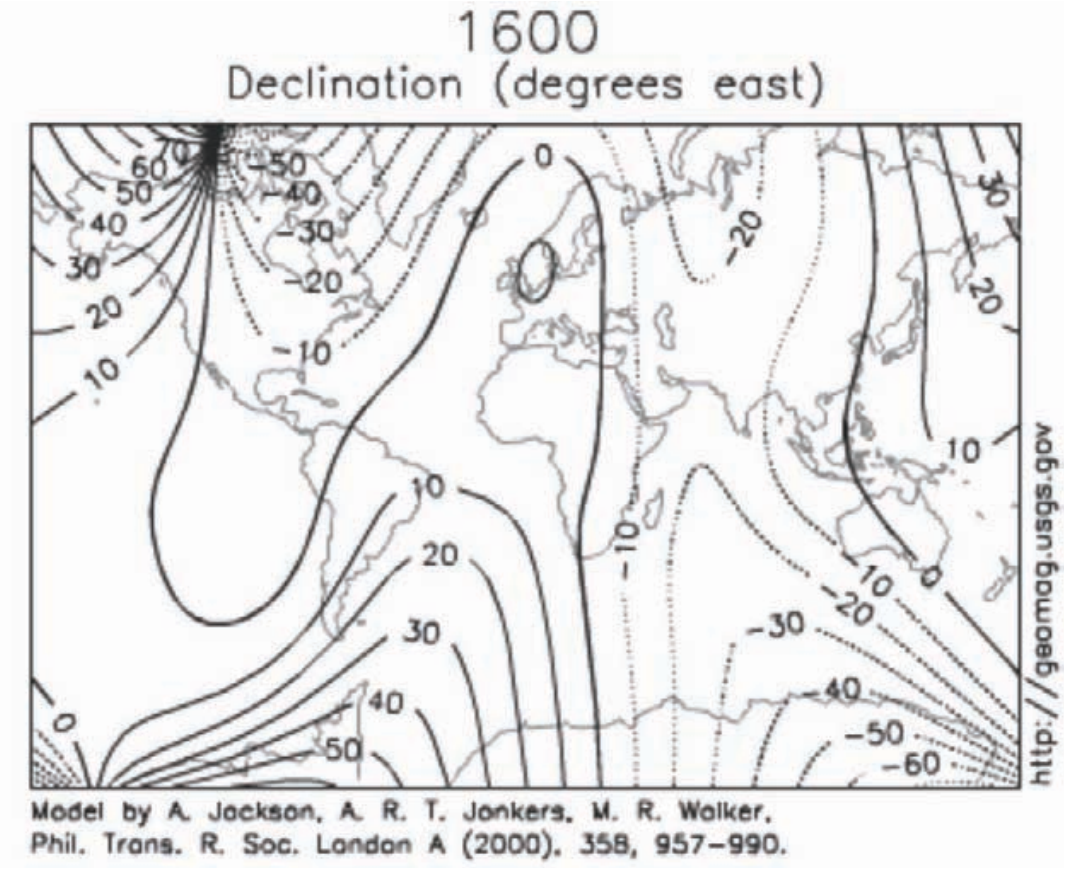

**Figure 13**: Declination of the Earth's magnetic field in 1600. (U.S. Geological Survey data, via Wikimedia Commons; public domain)

Confronted by the potential difficulties imposed by adverse weather conditions and the prevailing ocean currents, Ayanz eventually abandoned his engineering solution. Instead, he began to look into using the magnetic compass for the same purpose—clearly jumping onto the bandwagon led by Arias de Loyola and Fonseca Coutinho. He carefully studied the latter's proposed solution but quickly reached the conclusion that Fonseca Coutinho's suggestion of a fixed compass needle that always pointed to true North was a wholly impractical proposition, given small- and large-scale variations in the Earth's magnetic field. (Figure 13 shows the declination of the Earth's magnetic field in 1600, exhibiting clear deviations from straight-up North–South meridians.) Ayanz demonstrated the impossibility of Fonseca Countinho's solution in an undated memorandum[15] and an accompanying table, which were sent to the *Conseja de Indias* and the king on 13 November 1610 (Fernández de Navarette, 1852: 138–139; García Tapia, 2016).

A mere two weeks later, on 26 November 1610, Admiral Diego Brochero de la Paz y Anaya was tasked with the verification of Ayanz' novel compass needle solution (García Tapia, 2016)—involving a comparison of the behaviour of fixed and variable needles. Nevertheless, the *Conseja* decreed that Ayanz would not be heard until the secrets of Fonseca Coutinho's method had been disclosed. Moreover, the solution presented by Arias de Loyola was to be preferred over Ayanz' approach. Unfortunately, the technical details of Ayanz' proposed longitude solution are lost in the depths of history.

**6 GALILEO'S SOLUTION**

Galileo Galilei [16] was undoubtedly the most illustrious of the Spanish longitude prize contenders. Yet despite his heralded stature, it is perhaps surprising that his proposal did not manage to enthuse the Spanish monarch.

Shortly after sunset on 7 January 1610, Galileo busied himself with testing a telescope prototype recently manufactured to his own design. He decided to point the telescope at Jupiter and noted,

> … when I was viewing the constellations of the heavens through a spyglass,[17] the planet Jupiter presented itself to my view, and as I had prepared for myself a very excellent instrument, I noticed a circumstance which I had never been able to notice before, owing to want of power in my other spyglass, namely, that three little stars, small but very bright, were near the planet; and although I believed them to belong to the number of the fixed stars, yet they made me somewhat wonder, because they seemed to be arranged exactly in a straight line, parallel to the ecliptic and to be brighter than the rest of the stars equal to them in magnitude. (Galilei, 1610)

The next day,

> … when on 8 January, led by some fate, I turned again to look at the same part of the heavens, I found a very different state of things, for there were three little stars all west of Jupiter, and nearer together than on the previous night, ... yet my surprise began to be excited, how Jupiter could one day be found to the east of all the aforesaid fixed stars when the day before it had been west of two of them; and forthwith I became afraid lest the planet might have moved differently from the calculation of astronomers, and so had passed those stars by its own proper motion.

Given the state of astronomical knowledge at the onset of the seventeenth century, it is no surprise that Galileo was perplexed by his observations. If the bright ‘stars’ he had seen near Jupiter were indeed members of the ‘fixed’ (background) stars, it transpired that Jupiter must have moved eastwards overnight. Yet, it was common knowledge among astronomers at the time that, like all other known planets in our solar system, Jupiter was bound to move in the opposite direction to that traced by the fixed stars. He grew more anxious on 9 January, a cloudy night, and on 10 January 1610, only two of the three ‘stars’ were visible through his telescope. A fourth star appeared on 13 January.

Repeat observations over the next few weeks showed that the four stars closely followed Jupiter in its orbit with respect to the background of fixed stars, merely changing their relative positions with respect to the planet and each other over time. Galileo’s thus concluded that the four stars were celestial bodies in orbit around Jupiter:

> I have discovered Four Erratic Stars, neither known nor observed by any one of the astronomers before my time, which have their revolutions round a certain bright star [Jupiter], one of those previously known, like Venus and Mercury round the Sun, and are sometimes in front of it, sometimes behind it, though they never depart from it beyond certain limits. (Galilei, 1610)

Galileo referred to his ‘Four Erratic Stars’ as the ‘Medicean Planets’, in honour of his patron, Cosimo II de’Medici, Grand Duke of Tuscany. He referred to the individual moons by Roman numerals I–IV, increasing with distance from Jupiter. The current names of these Galilean moons are Io, Europe, Ganymede and Callisto, as initially suggested by Johannes Kepler.

The first attempts to characterise the orbits of Jupiter's Galilean moons, with orbital periods between 1¾ and 17 days, for their specific use as a celestial clock, indicating ‘astronomical time’ through observations of the moons’ ephemerides, commenced almost immediately. This attests to the efficiency of the dissemination of Galileo’s novel insights across the continent. Johannes Kepler in Prague, Thomas Harriet in Syon (near London), Jesuit mathematicians at the *Collegio Romano* in Rome and, most likely, Simon Marius in Ansbach (near Nuremberg, Germany) proceeded to observe the movements of Jupiter’s satellites by the end of 1610. In addition, the French astronomer Nicolas-Claude Fabri de Peiresc observed the Galilean satellites from Aix-en-Provence between November 1610 and June 1612. His associate, Joseph Gaultier de la Valette, prepared the first observational tables. Assuming that the moons’ orbits were circular and confined to the ecliptic plane, Peiresc subsequently prepared an almanac containing the corresponding ephemeris tables.

Galileo's own annotations indicate that, as early as 1612, he was also engaged in observing the eclipses of Jupiter's moons. By September 1612, he had calculated accurate periods and made ephemeris tables of their projected positions. These were sufficiently accurate to predict their motions, eclipses (by Jupiter's shadow) and occultations (by the planet itself) several months ahead (O'Connor and Robertson, 1997). Jean Lombard, another of de Peiresc's associates, attempted to use the well-known ephemerides of the Galilean satellites for longitude determination across the Mediterranean as early as 1612. That year, he travelled to Marseille, Malta, Cyprus and Tripoli (Libya), observing whenever possible the arrangement of Jupiter's satellites. He intended to compare his observations with de Peiresc's tabulated ephemerides, but he quickly concluded that the positions of the satellites do not change sufficiently rapidly for this to become a viable approach.

Nevertheless, in 1612 Galileo himself fronted a first proposal to use the eclipses and occultations of Jupiter's moons to determine absolute time and, hence, provide a means to determine one's longitude, at sea or on land. Galileo's proposal was part of a more comprehensive collaboration proposal made to the Spanish Crown by Cosimo II de'Medici. The proposal was rejected, and rightly so, because of the inability to make precise observations with a telescope from a continuously pitching and rolling ship. (Land-based observations of Jupiter's moons were later used routinely to determine or confirm longitude measurements.) In addition, the eclipses of the Galilean moons were not quite instantaneous, thus rendering the determination of absolute times from any location on Earth impractical. Moreover, Galileo's telescope was not really appropriate for the observational challenge. His eyepiece consisted of a concave lens, resulting in an extremely small field of view on the sky. This would make it difficult to locate the planet and, more difficult still, to keep it in the field of view for the time necessary to observe the eclipses of its satellites.

Despite these objections, Galileo was confident that any of these difficulties could be overcome, thus redoubling his efforts to compile sufficiently accurate ephemeris tables. He also proposed to manufacture telescopes with magnifications of 40–50×, which he would be taking with him to Spain. At the same time, he would bring along an expert in their use so as to pass on his knowledge to the mariners ultimately tasked with operating the instrument. In addition, Galileo committed to preparing annual almanacs containing the ephemerides of Jupiter's satellites and to providing written instructions. He thus continued his lobbying forcefully, which culminated in a second submission to the *Casa de la Contratación* in 1616. Once again, Count Orso d'Elci advised Galileo of the Spanish Crown's rejection of his proposal:

> From your writing I understand that from the time difference in which the same aspect of these stars around Jupiter is observed, the difference in true longitudes of those cities or places can be quickly known. But for this it is mandatory and necessary to see the aforementioned stars and their aspects. I don't know how this can be done at sea, or at least as frequently and quickly as necessary for the person who sails. Because, leaving aside that telescopes cannot be used on ships due to their movement, even if they could be used they would not work during the day or at night with overcast weather, because the stars are not visible, and the navigator needs to know hour by hour his longitude … (Drinkwater Bethune, 1830)

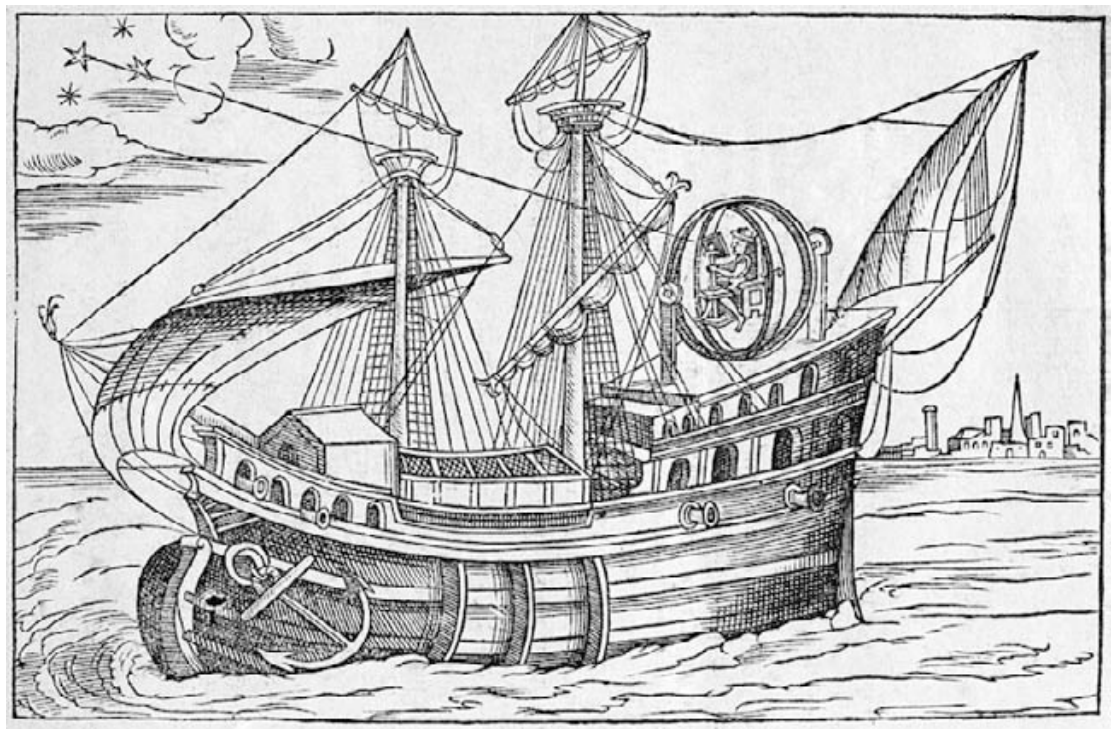

**Figure 14**: Possibly the first illustration of a gimballed (marine) chair on a ship, designed to allow astronomical observations from a stable position. (Besson, 1567–1569; public domain)

This series of events triggered a protracted, ultimately unsuccessful, 16-year-long correspondence between Galileo, the Tuscan ambassador at the Spanish court, the Spanish *chargé d'affaires* in Rome and the cosmographers representing the Spanish Crown (Howse, 1980; O'Connor and Robertson, 1997). Galileo even offered to travel to Spain and reside in Seville—in close proximity to the *Casa de la Contratación*—or in Lisbon (also part of King Philip III's empire), for as long as necessary (Barrado Navascués, 2015).

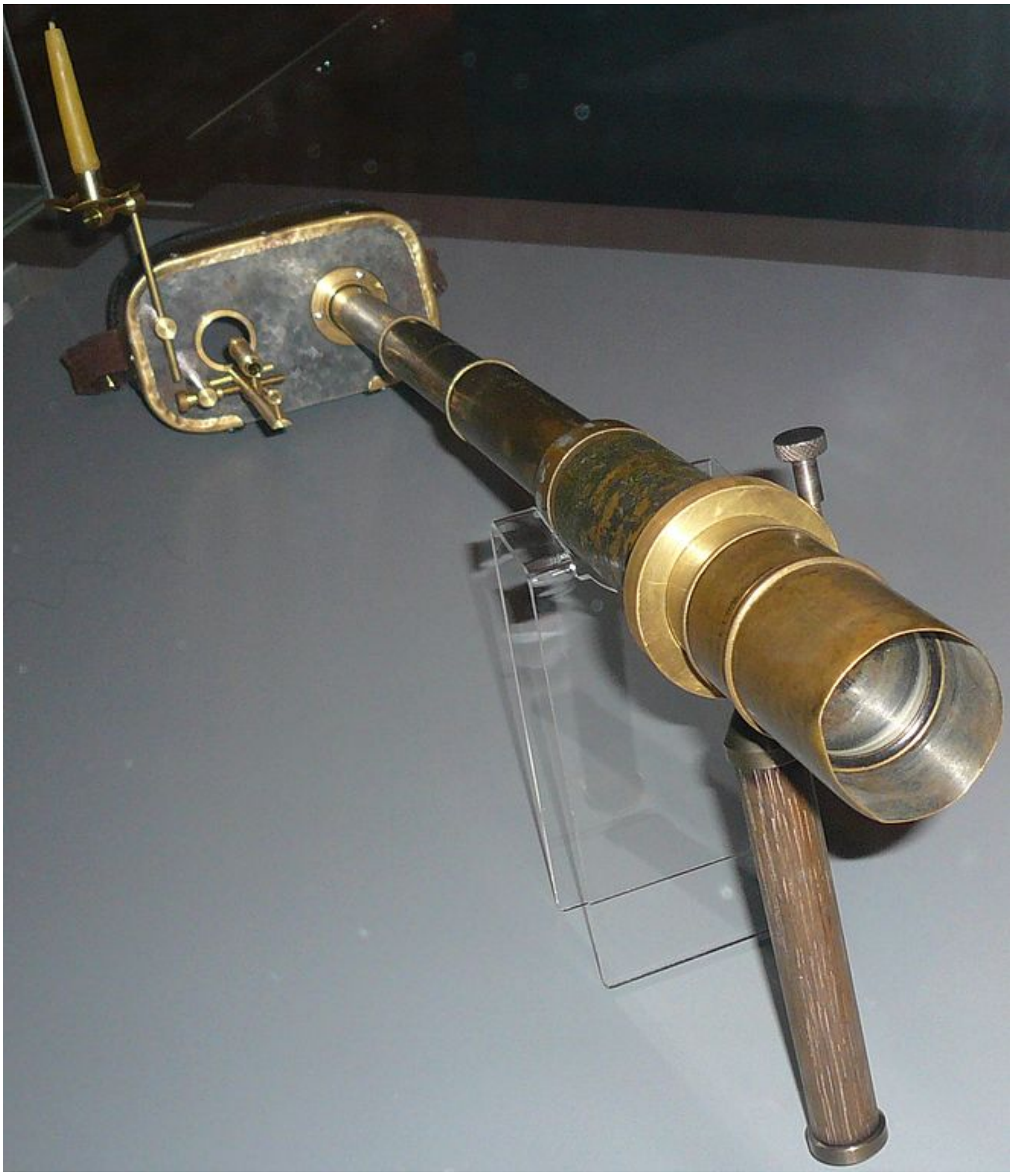

**Figure 15**: Reproduction of a *celatone* based on an interpretation of Parlour's (1824) 'apparatus to render a telescope manageable on shipboard', itself a reinvention of Galileo's *celatone*. (Museum at the Royal Observatory, Greenwich, UK; Creative Commons Attribution-Share Alike 4.0 International license)

Despite the second rejection, in September 1617 Galileo proceeded to test a new approach. He approximated a stable, gimballed suspension by positioning a chair in a small boat that was floating in a pool of water on the deck of a ship in the Tuscan port of Livorno (see Figure 14). On his head, he was wearing a *celatone*, a helmet with a telescope mounted to its eye slit (an example is shown in Figure 15), which he had personally constructed in the Grand Duke's workshop. Its viewfinder could be adjusted to align the axis of the telescope with the observer's eye, allowing him to follow Jupiter's moons, while the other (unaided) eye could locate Jupiter itself. In his prototype *celatone*, he had mounted a low-magnification telescope inappropriate for adequate observations of Jupiter, but his main aim was to show that a telescope could be used on ships—to unexpected acclaim by the military commander Giovanni de'Medici. Nevertheless, Galileo conceded that even on land one's heart rate could cause Jupiter to rhythmically jump out of the telescope's field of view (de Grijs, 2017: Ch. 3). A few months after Galileo's first attempt, he had his friend and student Benedetto Castelli test it at sea. Once again, Galileo wrote to Count d'Elci,

> This is an art still in development, based on new principles and methods, that needs to be wrapped, cultivated and fostered so that with practice and time the fruits will be obtained since it already contains the seeds and roots, but again, he did not manage to convince the Spanish Crown of his method's viability. Ultimately, none of Galileo's proposals were honoured by the *Casa de la Contratación.*

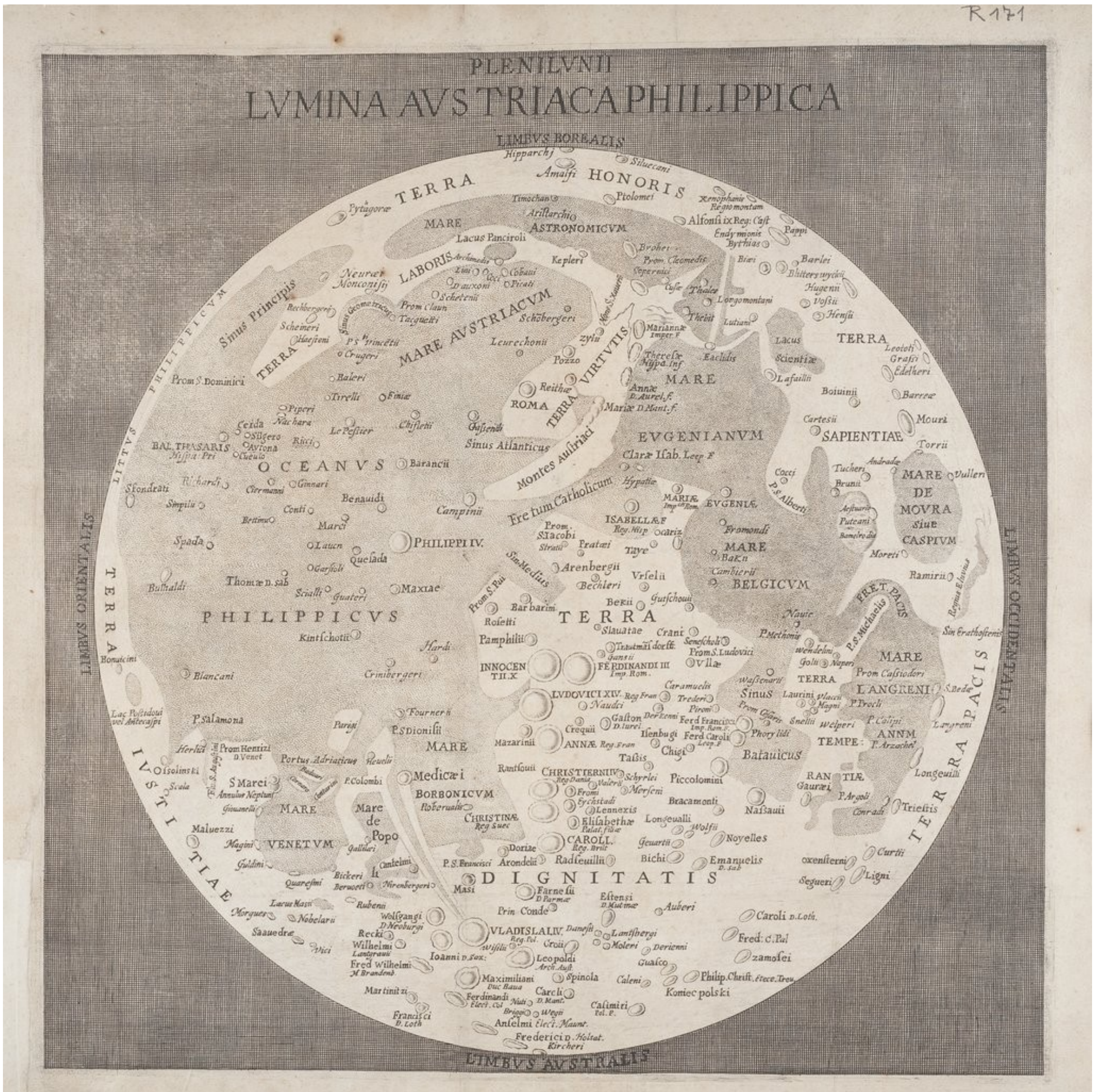

**Figure 16**: Van Langren's celebrated map of the Moon, *Plenilunii lumina Austriaca Philippica* (1645). (Bibliothèque nationale de France; public domain)

## 7 MICHAEL FLORENT VAN LANGREN

The other outside contender for the Spanish longitude prize was Michael Florent van Langren, a Flemish cartographer, astronomer and mathematician, who eventually succeeded his father as cosmographer and mathematician at the Spanish court. Van Langren is best known as author of the *Plenilunii Lumina Austriaca Philippica* (1645; see Figure 16), the first published map of the Moon with lunar features based on the names of kings, princes, politicians, scientists, explorers, religious leaders and saints (Whitaker, 1999). He drew the lunar craters as if they were illuminated by the morning Sun, a technique still in use today (Navarro Brotons, 2018).

He created the first known graph of statistical data that showed a large range of previously measured longitude differences between Toledo (Spain) and Rome (Friendly et al., 2010). As one of his first scientific pursuits, he suggested that these estimates could be improved, most notably so at sea, by observing the phases of the Moon, and specifically by determining the absolute timing of the rising and setting of peaks and craters—at any time and not only during lunar eclipses as had been suggested previously.[18] Van Langren realised that as the lunar phases progress from new to full moon, the Sun progressively illuminates different lunar features from East to West; these same features disappear progressively from full to new moon. The approach he proposed was simple and, hence, potentially practically viable; it involved measuring astronomical times, which could then be compared with ephemeris tables. Having commenced his astronomical forays in 1621, by 1625 he was sufficiently convinced of his approach to determining longitude using the timings of appearing and disappearing lunar features that he confidently presented his method as a possible solution of the longitude problem to the *Infanta*[19] Isabella Clara Eugenia, daughter of Phillip II (Navarro Brotons, 2018):

> Most Serene Highness,
>
> Miguel Florencio van Langren, Mathematician to His Majesty, says that his Grandfather and his Father, Cosmographer to His Majesty, have been the first who have invented Globes for the direction of navigators, and the supplicant, emulating them has attained with great study and attention some fundamental and concealed aspects of the aforementioned art as well as others, and one of the main is that of Longitude, by which it is possible to lay out perfectly all the Terrestrial description, which has countless errors as can be seen in the writings of different authors, because comparing two maps or tables of longitude of different authors, by no means do they concur between them as Your Serene Highness will see in the following example. If the Longitude between Toledo and Rome is not known with certainty, consider Your Highness, what it will be for the Western and Oriental Indies, that in comparison the former distance is almost nothing. So to amend these deficiencies and to find the true distances of the towns of the Earth, it would be necessary that Your Serene Highness be pleased to supplicate to His Majesty to dispatch a Patent so that he can send his corresponding and printed instructions for all the Earth, both to the East and to the West, ordering in it that all interested in the art observe what the supplicant advises them, promising that many benefits will derive for navigation, and eternal memory for His Majesty and Your Highness, for having ordered this general correspondence of the art, and Your Most Serene Highness will receive it very particularly. (Friendly et al., 2010)

From a letter the *Infanta* Isabella sent to King Philip IV[20] in 1625, we learn that van Langren also informed her about a second method he had conceived, but which is only available in the form of an as-yet-unsolved cipher (van Langren, 1644). *Infanta* Isabella engaged Erycius Puteanus and Godefroy Wendelin, scholars from the Spanish Netherlands,[21] to examine van Langren's proposed solution. Van Langren demonstrated the details of his method to these eminent scholars on 5 March 1631, following a delay that had been incurred on account of van Langren's earlier, existing travel commitments to Spain. In his demonstration, he outlined both his main method as well as two additional methods using nighttime lunar observations at the meridian or at any azimuth, of which however no records survive (van Langren, 1644). Van Langren had obtained such observations from both Madrid and Brussels.

Upon receiving the experts' endorsement, the *Infanta* Isabella dispatched van Langren back to Madrid, carrying recommendations to her nephew, King Philip IV. The king took a keen interest in van Langren's work and asked him to observe the sky and the Moon with him through his telescope. As we saw already, King Philip IV subsequently ordered van Langren's observations to be published, at royal expense, in the form of the *Plenilunii Lumina Austriaca Philippica*.

By royal decree, issued by the *Conseja de Indias*, van Langren was requested to compose a set of instructions regarding the theoretical and practical applications of his method so as for the pilots of the *Casa de la Contratación* to calculate their longitude while sailing at a given, constant latitude. He argued that his method was much more precise than anything proposed previously, with associated errors of less than two or three leagues at mid-latitudes, increasing to less than four or five leagues at the Equator.

Van Langren formally submitted a proposal outlining his new method on 7 January 1632. His treatise containing the method's technical details reached the *Conseja de Indias* on 10 May 1633. Upon their technical assessment, two eminent mathematicians, the Marquez de Oropesa and Lorenzo Ramírez de Prado, advised the king to award the astronomer 4,000 ducats (van Langren, 1644). Van Langren himself was convinced of the merits of his method, however, and asked for more:

> And in this regard His Majesty offered to the Inventor of such solution great rewards, and in particular to Luis Fonseca 6,000 ducats every year, and then to Juan Arias, 2,000 ducats every year for a lifetime; so if his Majesty sends to this Supplicant the assurance of a prize that his Royal Highness judges appropriate, [the Supplicant] will report the aforementioned secret to His Majesty, because finding this invention and not getting any reward would be honourless.
>
> [He, that is, van Langren] also supplicates that His Majesty shelters him against the objections that some could put, saying that my invention is old and known and that I should not interfere in this, as I think that it must be sufficient if it is good and useful. (Friendly et al., 2010)

However, Fernando de Contreras, secretary of the *Conseja de Indias*, was not as favourably disposed towards the Flemish astronomer. Therefore, and despite several additional endorsements, including by the Flemish mathematician Jean Charles della Faille (van der Viver, 1977), no award was issued and van Langren hence declined to explain his method in detail.

In 1634, van Langren returned to Flanders by order of the king, without any tangible prospects for a reward. Prior to his departure, he provided full details of his main proposal, *Advertencias de matemático de su majestad, a todos los profesores y amadores de la matemática tocantes a la proposición de la longitud por mar y tierra, que ha hecho a su majestad católica* (Madrid, 1634), leaving della Faille to further distribute information about his method and represent his interests at the Spanish court (Navarro Brotons, 2018). A decade hence, he published a second summary of his method to determine longitude, again in Spanish, *La verdadera longitud por mar y tierra: demostrada y dedicada a Su Majestad Católica Philipo IV [...] Con las censuras y pareceres de algunos renombrados y famosos mathematicos deste siglo [...]* (Brussels, 1644).

Ultimately, van Langren's approach to longitude determination failed, most importantly because lunar features appear gradually rather than instantaneously.

## 8 STRAGGLING INNOVATORS

In his disclosure of 1644, van Langren mentions that in 1629 Christoforo Borri, an Italian-born Jesuit whom van Langren referred to as Cristóval de Bruno Milanez, presented an "invention to navigate from West to East, based on vacillations of the magnetic needle and tracing the respective longitudes through points of equal declination" (Andrade Corvo, 1882). This was apparently inspired by the Portuguese cosmographer Valentim de Saa. Also known by his Portuguese name Cristóvão Bruno, the Jesuit had developed his method following extensive travel by sea in the Far East, from his native Milan to Macao and Cocinc(h)ina (present-day Vietnam). Although the details of this scheme have been lost, it is said to have been submitted in both writing and graphical form—perhaps one of the first isogonic maps ever made, showing the spots where a magnetic needle makes the same angles with the meridian (Kircher, 1641; Jonkers, 2003).

In a letter dated 17 March 1629, Bruno triumphantly exclaimed,

> My business of the invention from the West to the East has already been examined, and approved in this Royal Council of Portugal, where all the sages and intellectuals [skilled] in this issue from the entire kingdom assemble together with the Pilots. And later it was approved by the Council in Madrid. Finally, the King commanded that this March and armada of three ships and six galleons under the Viceroy would be dispatched to India navigated by this my invention. Necessary instruments for all the ships have already been made at royal expense, and the Pilots have been instructed and obligated to comply with the invention, … (Dror and Taylor, 2006: 46)

Next, in December of that year, he announced that …

> The King ordered that there would be the last council to determine the prize; and in this he ordered to give me maintenance and the cost of the publication of the book on this issue, also the King ordered to tell me he wanted to see the invention. Two Crowns, of Portugal and Castille, were fighting over me: this one wanted me to return to Lisbon to supervise their navigation; that one wanted me [to go] to Seville in order to apply the same invention to their Navy of the West Indies, including the Philippines, … (Dror and Taylor, 2006: 47)

His initial success in convincing the king of his method's promise caused him to adopt a brazen and victorious attitude, gaining him a reputation of being presumptuous and bizarre, which eventually led to his downfall. Despite his scheming to play off the Spanish and Portuguese Crowns against each other, he never received the 50,000 cruzados he had initially been promised by the Spanish Crown.

Meanwhile, while employed as a teacher at the *Colégio de Santo Antão o Novo* in Lisbon, Bruno had written a manuscript on the art of navigation, *Arte de navegar* (Lisbon, 1628). He wrote the manuscript in Portuguese, and so in order to make it more generally accessible, Father Lejeunehomme was tasked with its translation into Latin:

> … our fathers encouraged me to turn this book into Latin, because everybody thinks that this book would have a wider distribution than in Portugal; perhaps I will be occupied with that this year. (Dror and Taylor, 2006: 46)

Commenting on his work, Lejeunehomme noted that Bruno "found a means to identify the distances of longitude from East to West[22] and a new way for better navigation, which is in grand vogue here [in Portugal]." In fact, Bruno conceded in his manuscript that he was not referring to an "invention but *aperfeiçoamento* [improvement] of the old idea of using a clock to determine longitude" (Sommervogel, 1885). His clock, an hourglass that would last for at least six hours and possibly longer, should be equipped with the intervals between the hour indicators divided into 15 parts, representing the 15 degrees of the Earth's rotation per hour. While valid theoretically, this idea fell short of expectations in practice. It would take until well into the eighteenth century before a viable marine timepiece was manufactured.

In 1630, shortly after Bruno's failed attempt at securing the Spanish longitude prize, one Antonio Ricci from Genoa showed up and laid claim to a share of the prize on the basis of longitude measurements he had obtained without even having observed the sky. However, beyond van Langren's original 1644 record of Ricci's appearance, and a number of superficial historical references to the same event (e.g., Lelewel, 1852), no other records of Ricci's submission survive.

Around the same time, the Spanish philosopher and mathematician Juan Caramuel y Lobkowitz presented an altogether novel and systematic approach to longitude determination to the *Conseja de Indias*, one based on lunar positions. A Cistercian ecclesiastical scholar from Madrid, sometimes referred to as 'Spain's Leibniz', Caramuel has been celebrated as "the most fertile, erudite polymath of his century" (Menéndez Pelayo, 1883–1889). Yet, methodologically, apparently "he was more ingenious than judicious, more marvelous than practical" (Mayans, 1992).

Van Langren (1644) comments that the theologian asked for an award of 10,000 ducats in return for his disclosure. However, his proposed solution to the longitude problem was not his own to take credit for. In 1625, Caramuel joined the Cistercian Order at the Monastery of La Espina (García Camarero, 2018), in the diocese of Palencia (north of Valladolid), where Pedro de Ureña (de Hereña[23]) took him under his wings. Blind from birth, de Ureña was nevertheless a formidable mathematician and astronomer. By 1615, he had devised a method for longitude determination based on lunar motions, laid out in his treatise *De Astronomía*. The manuscript was ready and fully licensed for printing already by 1620, but de Ureña passed away before he saw his project through to fruition (Sanhuesa Fonseca, 1999; de Barcelos e Coles, 2014). He left the dissemination of his masterwork to Caramuel instead (Velarde Lombraña, 1982).

For the first time, a novel method proposed to the *Conseja de Indias* led to positive and promising results, in turn demoting magnetic declination approaches to find *el punto fijo* to little more than interesting historical side notes (Velarde Lombraña, 1985). Despite his success, it took Caramuel until 1670 before he managed to get the details of the new method printed as part of his treatise *Mathesis Biceps: Vetus Et Nova*:

> [Special Geometry; Corollary II, Lemma 19]
> It is manifest from the following passages, that we must remain in the exact same locations to be able to attain to knowledge about the Longitude; if at these locations, of which the Longitude of one is investigated, these two components must be maintained, to wit, the Moon at any azimuth, and the altitude of any of the Stars, at the same moment in which the observation is made. Determination of the Longitude under investigation is much easier, and more so in its explanation, by means of Lunar Eclipses: but they rarely occur and are not always well visible. (Caramuel y Lobkowitz, 1670)

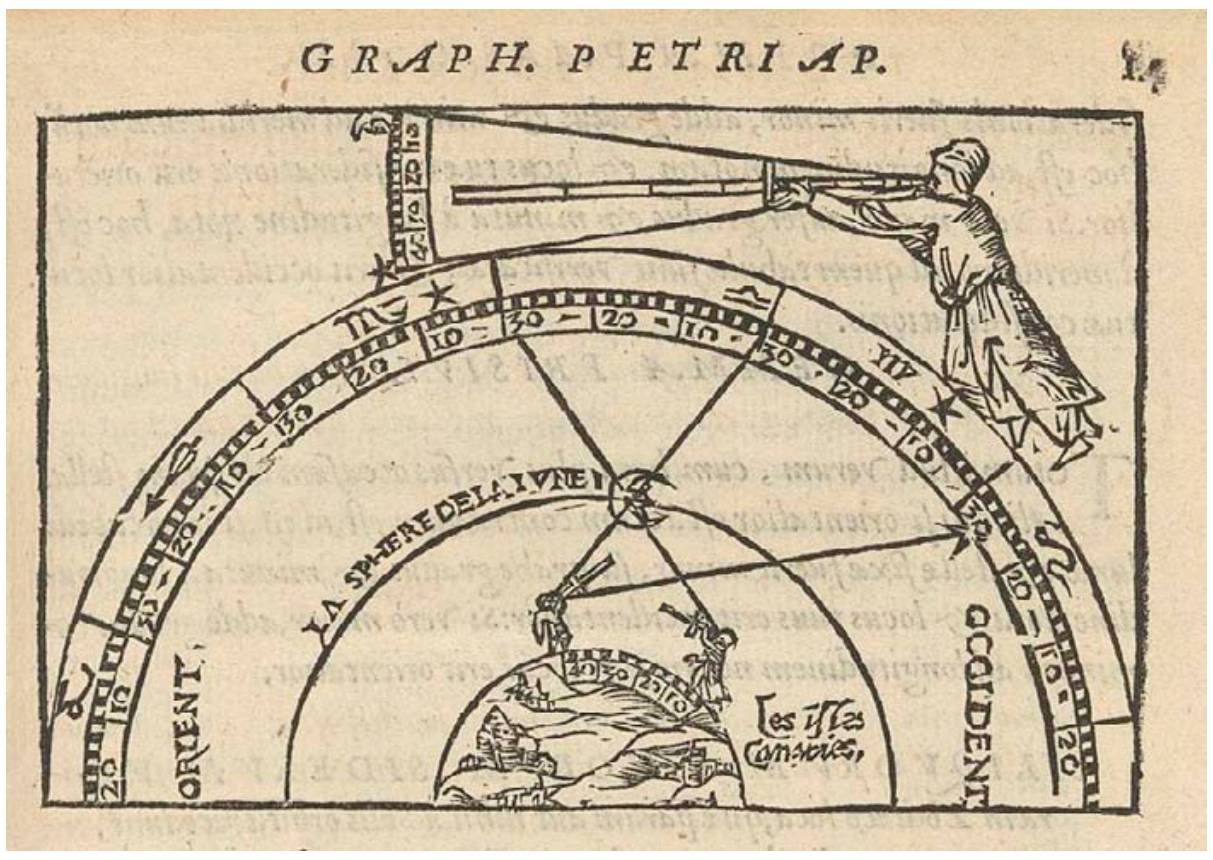

**Figure 17**: Petrus Apianus's depiction of the principles underlying the lunar distance method, published in his treatise *Cosmographia* (Landshut, 1524). (Wellcome Trust; Creative Commons Attribution 4.0 International license)

From this introduction to the detailed method, it becomes clear that Caramuel's benefactor, de Ureña, had stumbled upon a method that soon became commonplace and would be known as the lunar distance method (for a recent review, see de Grijs, 2020; see also Figure 17): in principle, one can use the projected position of the Moon with respect to a set of reference stars to determine one's longitude.[24] Given a sufficiently accurate representation of the Moon's path across the sky, one can calculate such lunar distances, or 'lunars', for any location on Earth and for specific times. At any other position, one can then determine the local time(s) at which specific lunars occur. The difference between the tabulated time and the local time at the location of the observer can directly be converted into a longitude difference, provided that one can also adequately correct for the effects of parallax and atmospheric refraction (for full details, see de Grijs, 2020).

Meanwhile, in 1635 Caramuel built two towers, in Bruges and Dunkerke, both in the Low Countries, to help him better measure the Earth's shape and curvature. He employed a simple trigonometric parallax method, using triangulation and observers located at both locations. The towers at Bruges and Dunkerke were separated by 75 km, and both observers would compare their observations of the same celestial body to calculate that specific distance (Navarro Morales, 2011). Caramuel's prowess in mathematics, geography and architecture had made him a force to reckon with.

## 9 THE END OF AN ERA

Despite the urgency to find a solution to the longitude problem and the flurry of activity triggered by the prospects of significant prize money, King Philip III was not particularly enamoured by any of the schemes proposed (O'Connor and Robertson, 1997). As a consequence, no prize money was ever awarded beyond the expenses component (Gould, 1923; Jonkers, 2007). This led to a declining interest in the Spanish longitude prize from the mid-1630s onwards, with international efforts shifting to northern Europe.

The last credible proposal to the *Conseja de Indias* was submitted by Joseppe (Josepage) de Moro in 1637 (van Langren, 1644). De Moro had twice sailed around the world, and so he felt vindicated that a method based on magnetic declinations was the most promising approach. Despite the numerous failures of such schemes proposed previously, many meetings involving the relevant authorities and expert peer reviewers were called with renewed interest, even prompting the king to offer great riches if de Moro's method eventually proved successful. Success was still a distant prospect, however, and so—unsurprisingly (van Langren, 1644)—once again no prize was awarded by the Spanish Crown.

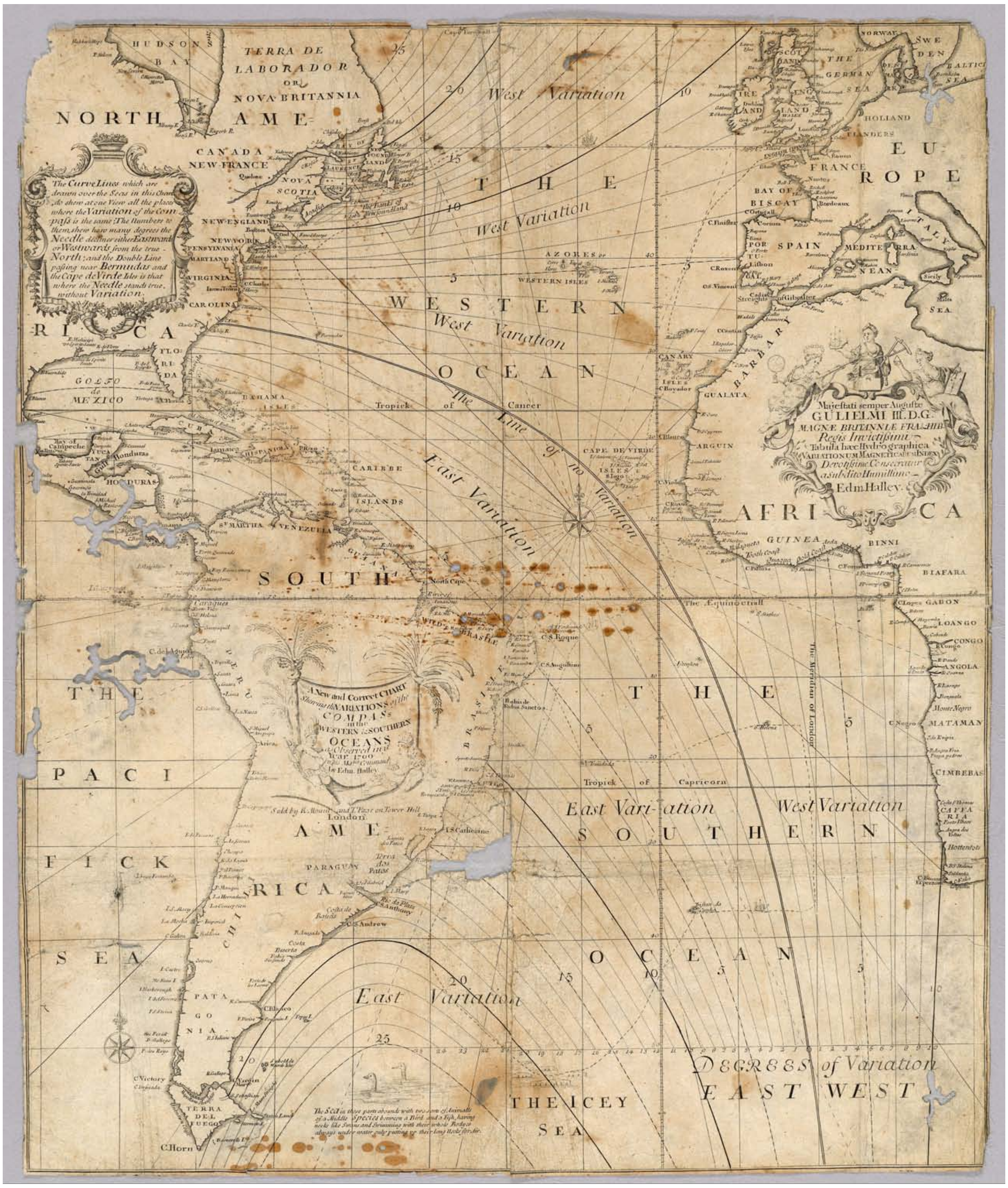

**Figure 18**: Halley’s first isogonic map, *General Chart of the Variation of the Compass* (1701). (Courtesy of the John Carter Brown Library at Brown University.)

The variety of responses to the Spanish longitude prizes offered a foretaste of what was to come a century later on a much larger scale, following the announcement of the British Longitude Prize in 1714 (e.g., de Grijs, 2017: Ch. 6). The enormous prize money on offer attracted a wide range of ‘projectors’, from genuine scientist-scholars to lunatics and those hoping to make a quick buck. This is reminiscent of the response elicited from de Cervantes, discussed in Section 4, and it is also reflected in the eighth and final panel of William Hogarth’s *The Rake’s Progress* (Barrett, 2013, 2016; de Grijs and Vuillermin, 2019).

However, despite the persistent noise introduced by unqualified projectors submitting outlandish proposals, genuine scientific advancement did result from these wide-ranging efforts, validated (or not!) by extensive sea trials. For one thing, magnetic declination solutions appeared doomed to fail, a valuable insight gained from these trials. Nevertheless, that disheartening prospect did not faze Edmond Halley, Britain's second Astronomer Royal, in pursuing precisely such a solution by the end of the seventeenth century (for a review, see de Grijs, 2017: Ch. 6). In 1683, Halley published his first of many scientific articles on the Earth's magnetic field. He devised a model of the Earth which, although later proven incorrect, implied that terrestrial magnetism had its origin deep in the planet's core—which we now believe to be correct. In 1696, he suggested that the Earth was composed of an outer shell surrounding an independently moving inner core. Both components would produce their own magnetic north and south poles, but the motion of the inner core was responsible for the observed behaviour of the terrestrial magnetic field.

In September 1699, Halley embarked on his second scientific voyage spanning the Atlantic Ocean, which lasted a year until 6 September 1700. During the voyage, Halley obtained extensive observations of the Earth's magnetic field. He published his results in the form of the *General Chart of the Variation of the Compass* (1701; see Figure 18), the first isogonic map of the North and South Atlantic Oceans, still used as an important reference today. Nevertheless, he was eventually forced to abandon his efforts to use magnetic-field measurements for longitude determination, since local variations from large-scale trends, combined with the inherent inaccuracies of contemporary compasses, rendered the approach unreliable.

Most of the other early solutions suggested to the *Conseja de Indias* were eventually validated and put to good use. Galileo's proposal to use the ephemerides of Jupiter's moons, while not viable on pitching and rolling ships, was employed extensively to obtain independent longitude determinations on land from the seventeenth century onwards, with a particular focus on South Atlantic shores. The lunar distance method of de Ureña and Caramuel quickly found its rightful place in the history of longitude determination, eventually becoming a mainstay before reliable chronometers became commonly available (e.g., de Grijs, 2020). And early suggestions to employ marine timepieces to crack the longitude problem were eventually validated by the development of John Harrison's marine watches, H1 to H4, by the end of the eighteenth century.

Early efforts in the vast Spanish empire, initially undertaken for political reasons, were thus eventually placed in a history of science context. Unfortunately, most historians of science focus almost exclusively on the advances made by north and northwest European scientists, often glossing over the seminal contributions made in response to the Spanish longitude prize announcements. This may well be due to the lack of comprehensive reviews of such Spanish contributions in English; even in Spanish and/or Portuguese, the literature highlighting early efforts of longitude determination in the Spanish empire are scarce—with the exception of the excellent yet largely unavailable exposition by Fernández de Navarette (1852). I sincerely hope that the present review may serve at least in part to return the early contenders for the Spanish longitude prizes to their rightful place in the history of science.

## 10 NOTES

[1] Figure 1 is a portrait gallery of the main characters driving the developments described in this article, if and when their images were available in the public domain. I encourage the reader to refer to this compilation figure whenever a new character voicing an original solution is introduced in a more than cursory manner.

[2] In 1469, the kingdoms of Aragon and Castile were united by the marriage of Ferdinand II of Aragon and Isabella I of Castile. This union laid the foundations for modern Spain. For brevity and simplicity, in this article I refer to 'Spain' to provide a geographic distinction from 'Portugal'.

[3] Like Figure 1, Figure 2 is a portrait gallery of the royalty and popes who played important roles in my narrative.

[4] Also known as the Peace of Alcáçovas–Toledo.

[5] One Italian league is equivalent to 2.67 nautical miles (4.94 km); one Portuguese league covers 3.2 nautical miles (5.9 km).

[6] Figure 6, like Figures 1 and 2, is a portrait gallery of officials, cosmographers and other important service personnel who were instrumental in driving the longitude discussion forward at this time.

[7] Also known as the Capitulation of Zaragoza.
[8] Note that Remando de los Ríos, who had gone to the Philippines in 1588, already claimed at that time to have determined his longitude by the declination of the compass needle (Vicente Maroto, 2001).
[9] The concept of a 'prime mover' goes back to Aristotle's notion of a primary cause or 'mover' of all motion in the universe (see, e.g., Nielsen, 1971).
[10] In 1577, two lunar eclipses were due to occur. King Philip II hence decreed that measurements be taken to determine the extent of the Castilian territories.
[11] The term 'projector' is often used in derogatory fashion when referring to someone who has jumped onto the bandwagon of an ill-defined 'project'. This designation goes back at least as far as Daniel Defoe's 1697 contemptuous exclamation, "A meer Projector then is a Contemptible thing, driven by his own desperate Fortune to such a Streight, that he must be deliver'd by a Miracle, or Starve; and when he has beat his Brains for some such Miracle in vain, he finds no remedy but to paint up some Bauble or other, as Players make Puppets talk big, to show like a strange thing, and then cry it up for a New Invention."
[12] A copy is available in the Navarrete collection in the Museo Naval in Madrid.
[13] The idea behind the *aguja fija*, the 'fixed needle', was that the compass needle was thought to point in the direction of magnetic North irrespective of one's geographic position.
[14] *Cédula de Privilegio*, Archivo General de Simancas, Chamber of Castilla section, inventory no. 174, leg. 262, 267 (440 pp.).
[15] Possibly of 29 October 1610 (Fernández de Navarette, 1852: 138–139).
[16] Galileo Galilei is usually simply referred to by his first name, a convention I have adopted here too.
[17] In 1610, the word 'telescope' had not yet been invented. After Galileo's rise to stardom on account of his treatise *Siderius Nuncius*, the Italian astronomer was invited as guest of honour at a banquet held on 14 April 1611 in Rome, hosted by Prince Frederico Cesi, founder and president of the Lincean Academy (Galileo was made a member 11 days later). There, Prince Cesi first referred to the Dutch spyglass at the basis of Galileo's newly acquired fame as a *τελεσκοπος*, a term coined by one of his other guests, John Demisiani, the Greek court mathematician of the Duke of Gonzaga. The new name was Latinized (and used in English from 1619) as *telescopium*, while starting life as *telescopio* in Galileo's Florentine Tuscan. The English form 'telescope' first appeared in 1650.
[18] This idea was also championed by the English astronomer and mathematician Lawrence Rooke (1622–1662), one of the founders of the Royal Society of London, who proposed to treat the Moon's irregular surface as gnomons on a sundial (Dictionary of National Biography, 1949–1950; see also Ward, 1740: 90–95; Ronan, 1960).
[19] The title *infanta* (princess) is given to the daughter of a reigning Spanish monarch who is not the heir-apparent. With her husband, Albert VII, Archduke of Austria, *Infanta* Isabela had been appointed joint sovereign of the Spanish Netherlands in 1601.
[20] Felipe Domingo Víctor de la Cruz had ascended the Spanish throne as King Philip IV on 31 March 1621. At the same time, he was made King Philip III of Portugal.
[21] The Habsburg Netherlands were ruled by the Spanish branch of the Habsburg family from 1556 to 1714. The region encompassed most of present-day Belgium and Luxembourg, parts of northern France, the southern Netherlands and western Germany. The *de facto* capital was Brussels.
[22] In fact, the method was explicitly suitable for determining longitudes from West to East, not from East to West (Dror and Taylor, 2006).
[23] Although van Langren's original manuscript spells the musician's surname as 'He**r**eña' (e.g. http://www.datavis.ca/gallery/langren/LaVerdadera.pdf), later renditions of van Langren's document invariably introduce a misspelling, rendering the name as 'He**rr**eña'.
[24] Note that calculated lunar positions usually refer to the Moon's centre coordinate, while tabulated lunars are based on distances to the lunar limb. One must, therefore, first apply a correction for the Moon's projected size on the sky (Dunlop and Shufeldt, 1972), which depends on its orbital position.

## 11 ACKNOWLEDGEMENTS

I thank Ángel R. López-Sánchez for having checked a number of my translations from Spanish to English.